\documentclass[lettersize,journal]{IEEEtran}
\usepackage{amsmath,amsfonts}
\usepackage{algorithmic}
\usepackage{algorithm}
\usepackage{array}
\usepackage{textcomp}
\usepackage{stfloats}
\usepackage{url}
\usepackage{verbatim}
\usepackage[dvipdfmx]{graphicx}
\usepackage{cite}
\usepackage{amsmath}
\usepackage{amssymb}
\usepackage{booktabs}
\usepackage{multirow}%
\usepackage{arydshln}%

\usepackage{subcaption}
\def\red#1{\textcolor{red}{#1}}
\def\blue#1{\textcolor{blue}{#1}}
\def\green#1{\textcolor{green}{#1}}

\usepackage[pagebackref=true,breaklinks=true,colorlinks,bookmarks=false]{hyperref}

\begin{document}

\title{Kernelized Back-Projection Networks\\ for Blind Super Resolution}

\author{Tomoki Yoshida${}^\dagger$, Yuki Kondo${}^\dagger$, Takahiro Maeda, Kazutoshi Akita, and Norimichi Ukita~\IEEEmembership{Member,~IEEE}
\thanks{All authors are with Toyota Technological Institute, Nagoya, 468-8511 Japan (e-mail: ukita@toyota-ti.ac.jp).\\{$\dagger$} The first two authors contributed equally to this work.\\{$\star$} : This code will be released in \url{https://github.com/Yuki-11/KBPN}.}}


\maketitle

\begin{abstract}
Since non-blind Super Resolution (SR) fails to super-resolve Low-Resolution (LR) images degraded arbitrarily, SR with the degradation model is required.
However, this paper reveals that non-blind SR trained simply with various blur kernels exhibits comparable performance as those with the degradation model for blind SR.
This result motivates us to revisit high-performance non-blind SR and extend it to blind SR with blur kernels.
This paper proposes two SR networks by integrating kernel estimation and SR branches in an iterative end-to-end manner.
In the first model, which is called the Kernel Conditioned Back-Projection Network (KCBPN), the low-dimensional kernel representations are estimated for conditioning the SR branch.
In our second model, the Kernelized Back-Projection Network (KBPN), a raw kernel is estimated and employed for modeling the image degradation.
The estimated kernel is employed not only for back-propagating a residual from its ground-truth but also for forward-propagating the residual to iterative stages.
This forward-propagation encourages these stages to learn a variety of different features in different stages by focusing on pixels with large residuals in each stage.
Experimental results validate the effectiveness of our proposed networks for kernel estimation and SR. We will release the code for this work${}^\star$.
\end{abstract}



\section{Introduction}
\label{section:introduction}

Image Super-Resolution (SR)~\cite{DBLP:conf/cvpr/TimofteGWG18} reconstructs a High-Resolution (HR) image,
$I^{HR}$, from its Low-Resolution (LR) image, $I^{LR}$.
We model the image degradation process from HR to LR as follows:
\begin{equation}
I^{LR} = (I^{HR} \ast K)\downarrow_{s},
\label{eq:down_model}
\end{equation}
where $K$, $\downarrow$, and $s$ denote a blur kernel, a downsampling
process, and a scaling factor, respectively.

\begin{figure}[t]
  \begin{center}
    \captionsetup[sub]{font=scriptsize, justification=centering}
    \begin{tabular}{c}
        
       \begin{minipage}{0.24\linewidth}
         \begin{center}
          \includegraphics[width=1\linewidth]{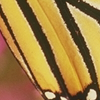}
          \subcaption{HR image\\
          ~}
          \vspace{0.7em}
          \label{fig:ppt3_hr}
        \end{center}
       \end{minipage}
       \begin{minipage}{0.24\linewidth}
        \begin{center}
          \includegraphics[width=1\linewidth]{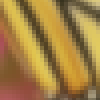}
          \subcaption{LR image\\ (PSNR=20.11)\\ }
          \vspace{0.7em}
          \label{fig:ppt3_only_blur}
        \end{center}
       \end{minipage}
       \begin{minipage}{0.24\linewidth}
        \begin{center}
          \includegraphics[width=1\linewidth]{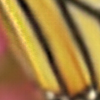}
          \subcaption{DBPN-Bl\\ (PSNR=24.90)}
          \vspace{0.7em}
          \label{fig:ppt3_dbpnbl}
        \end{center}
       \end{minipage}
       \begin{minipage}{0.24\linewidth}
        \begin{center}
          \includegraphics[width=1\linewidth]{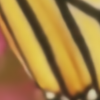}
          \subcaption{IKC~\cite{DBLP:conf/cvpr/GuLZD19}\\ (PSNR=27.64)}
          \vspace{0.7em}
          \label{fig:ppt3_IKC}
        \end{center} 
       \end{minipage}
       \\
       \begin{minipage}{0.24\linewidth}
        \begin{center}
          \includegraphics[width=1\linewidth]{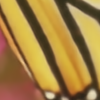}
          \subcaption{KOALAnet~\cite{DBLP:conf/cvpr/KimSK21}\\ (PSNR=25.46)}
          \label{fig:ppt3_koala}
        \end{center}
       \end{minipage}
       \begin{minipage}{0.24\linewidth}
        \begin{center}
          \includegraphics[width=1\linewidth]{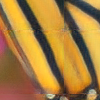}
          \subcaption{DASR--Gu~\cite{DBLP:conf/cvpr/WangWDX0AG21}\\ (PSNR=22.75)}
          \label{fig:ppt3_dasr_gu}
        \end{center} 
       \end{minipage}
       \begin{minipage}{0.24\linewidth}
        \begin{center}
          \includegraphics[width=1\linewidth]{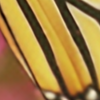}
          \subcaption{KCBPN\\ (PSNR=27.41)}
          \label{fig:ppt3_kcbpn}
        \end{center}
       \end{minipage}
       \begin{minipage}{0.24\linewidth}
        \begin{center}
          \includegraphics[width=1\linewidth]{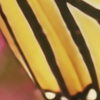}
          \subcaption{KBPN\\ (PSNR=\textbf{29.52})}
          \label{fig:ppt3_kbpn}
        \end{center} 
       \end{minipage}
      \end{tabular}
      \caption{
        Visual comparison of SR methods for an unknown kernel.
        The LR image~\subref{fig:ppt3_only_blur} is generated from its HR image~\subref{fig:ppt3_hr} by Eq.~(\ref{eq:down_model}) with the anisotropic Gaussian blur kernel~($\sigma_{x}/\sigma_{y}=2.6/4.0$).
        While the non-blind SR model trained with various blurs~\subref{fig:ppt3_dbpnbl} is comparable with blind SR~\subref{fig:ppt3_IKC} \subref{fig:ppt3_koala}, our method
        \subref{fig:ppt3_kbpn} reconstructs sharper boundaries.}
      \label{fig:ppt3_sample}
  \end{center}
  \vspace{-2em}
\end{figure}

\begin{figure*}[t]
  \begin{center}
    \begin{tabular}{c}
       \begin{minipage}{1\linewidth}
        \begin{minipage}{0.49\columnwidth}
         \centering
          \includegraphics[width=0.9\columnwidth]{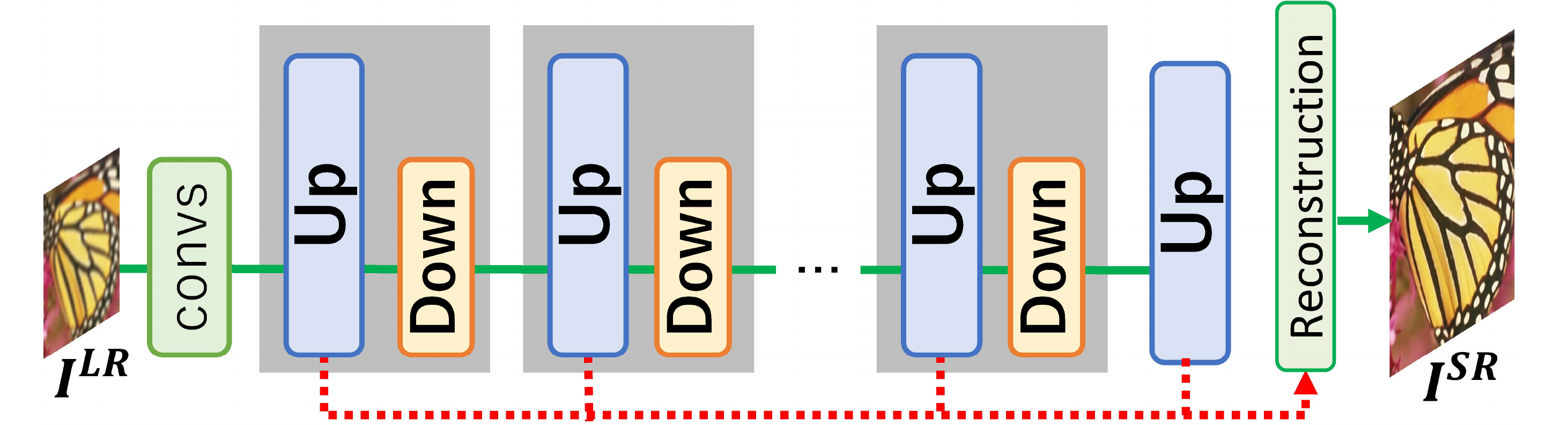}
          \subcaption{Iterative network for non-blind SR (e.g.,DBPN~\cite{haris2018deep})}
          \label{fig:dbpn}
        \end{minipage}
        \begin{minipage}{0.49\columnwidth}
         \centering
         \includegraphics[width=0.95\columnwidth]{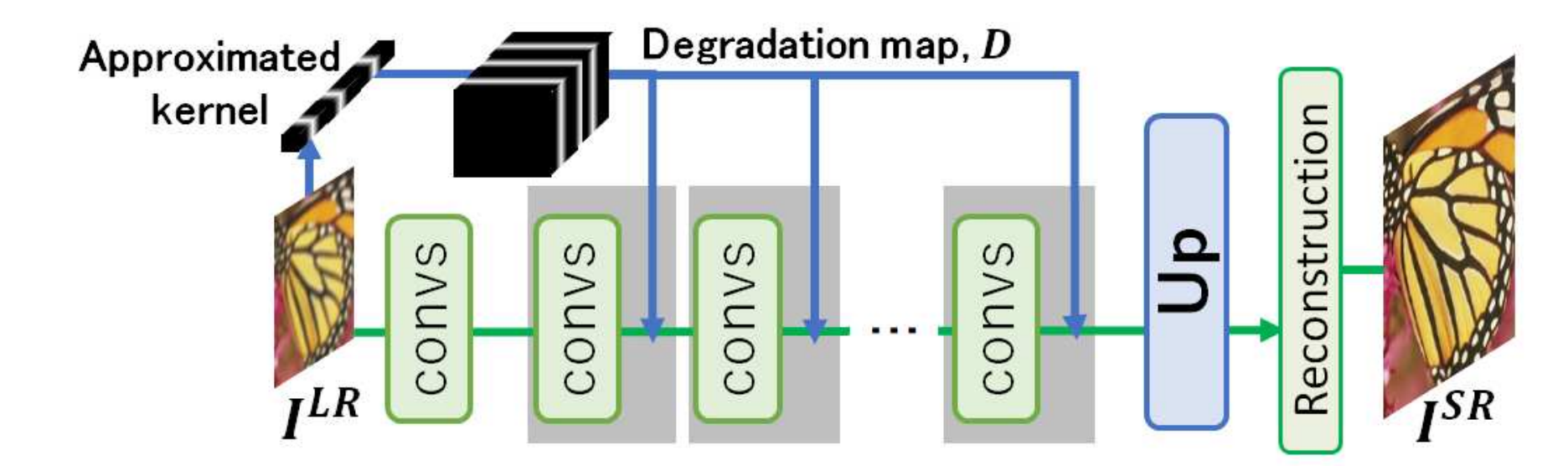}
         \subcaption{Iterative kernel estimation for blind SR (e.g., IKC~\cite{DBLP:conf/cvpr/GuLZD19})}
         \label{fig:ikc_sftmd}
        \end{minipage}\\
        \begin{minipage}{0.49\columnwidth}
         \centering
         \includegraphics[width=0.9\columnwidth]{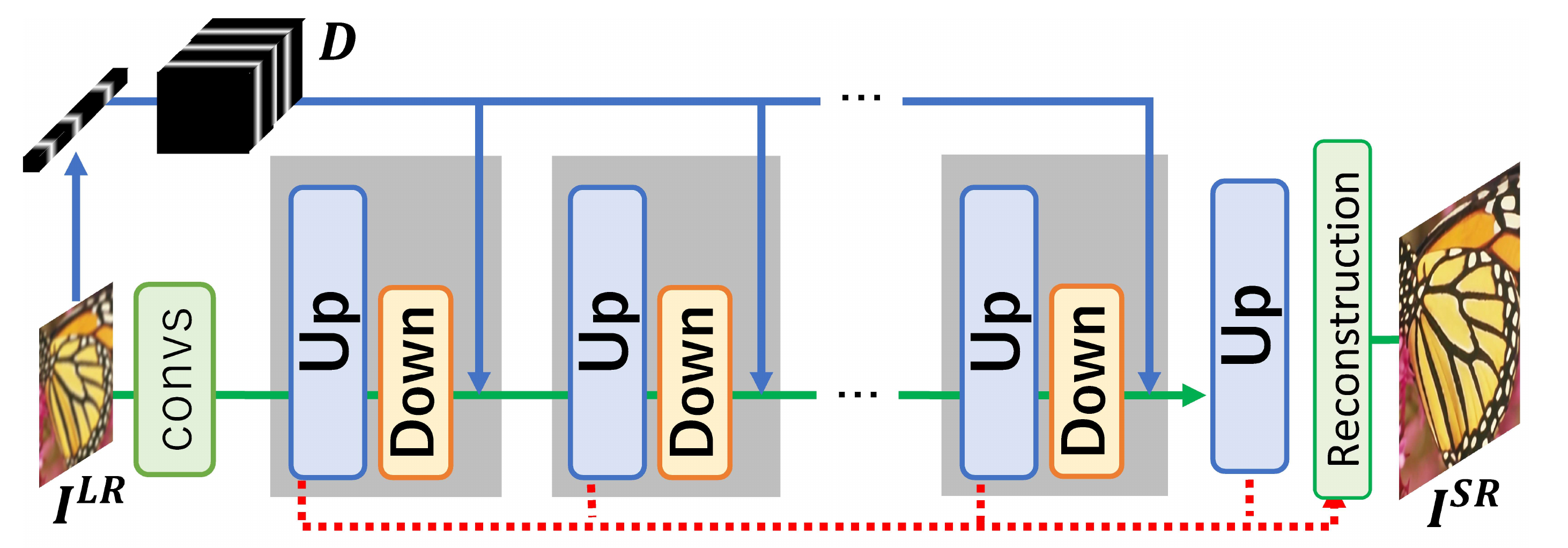}
         \subcaption{Proposed Method 1: KCBPN}
         \label{fig:kcbpn}
        \end{minipage}
        \begin{minipage}{0.49\columnwidth}
         \centering
         \includegraphics[width=0.9\columnwidth]{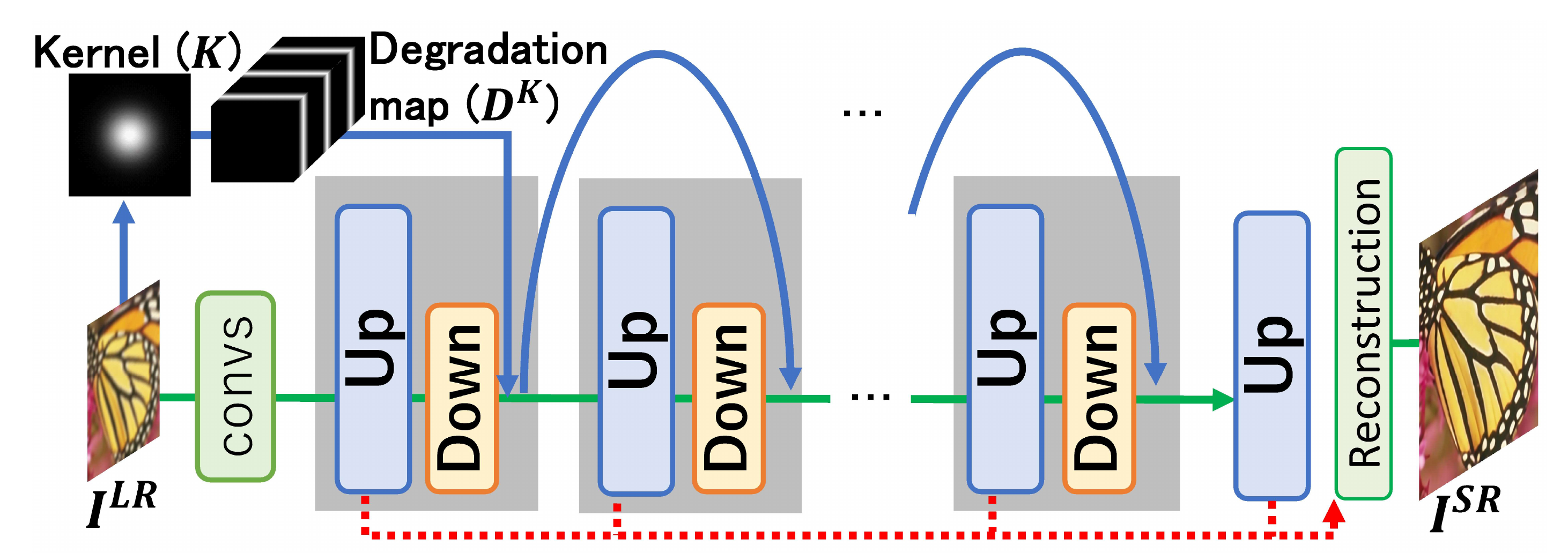}
         \subcaption{Proposed Method 2: KBPN}
         \label{fig:kbpn}
        \end{minipage}
       \end{minipage}
      \end{tabular}
      \caption{Comparison of iterative SR networks. Iterative
        stages are enclosed by gray rectangles. (a) Non-blind
        SR with iterative up- and
        down-projections~\cite{haris2018deep}. (b) Blind SR with
        kernel estimation~\cite{DBLP:conf/cvpr/GuLZD19}. (c) Our method 1 (KCBPN): Iterative
        up- and down-projections are integrated with kernel
        conditioning.
        (d) Our method 2 (KBPN): The image degradation process in Eq. (\ref{eq:down_model}) is explicitly modeled and employed for loss functions and SR feature enhancement.
        For highlighting the differences among (a), (b), and (c, d), several important details are omitted
        in this figure; see Fig.~\ref{fig:kbpn_stage_sr_kbpn2}, for the detail of KBPN.
      }
      \label{fig:compare_sr}
      \vspace{-0.5em}
  \end{center}
\end{figure*}

While early SR methods assume only a single degradation with no blur
kernel, recent methods model the blur kernel in
Eq. (\ref{eq:down_model})~\cite{DBLP:conf/iccv/RieglerSRB15,DBLP:conf/cvpr/ZhangZ018,DBLP:journals/tog/CornillereDWSS19,DBLP:conf/cvpr/GuLZD19,kernelgan,hussein2020correction,MZSR,DBLP:conf/cvpr/Liang0GGT21,DBLP:conf/cvpr/HuiLW021,DBLP:conf/cvpr/KimSK21,DBLP:conf/cvpr/WangWDX0AG21,MANet2021,DBLP:conf/nips/Luo0LWT20}.
While the superiority of the kernel SR methods is demonstrated in
comparison with those with no kernel, these non-kernel SR methods are trained with images degraded by only a single degradation process (e.g., bicubic downsampling with no blur kernel in Eq. (\ref{eq:down_model})) in the literature.
To validate the effectiveness of the kernel SR, it is worth exploring to revisit whether or not the non-kernel SR methods can be improved by being trained with variedly-degraded images.
Based on this revisit, this paper proposes two kernelized SR networks by integrating the advantages of kernel estimation and high-performance non-blind SR networks.
Our novel contributions are summarized as follows:
\begin{itemize}
\item We empirically found that non-kernel SR can be improved by being trained with variedly-degraded images.
\item Based on the above finding, two SR networks (Fig. \ref{fig:ppt3_sample} \subref{fig:ppt3_kcbpn} and \subref{fig:ppt3_kbpn} and Fig.~\ref{fig:compare_sr} \subref{fig:kcbpn} and \subref{fig:kbpn}) are proposed by iteratively integrating kernel estimation and SR feature enhancement.
  While the first network conditions SR features by a low-dimensional kernel representation, a raw blur kernel is employed with Eq.~(\ref{eq:down_model}) for our proposed SR feature enhancement in the second network.
\item Our SR feature enhancement is achieved with a residual between the input and estimated LR images in each iteration. Since the estimated LR image is affected not only by the reconstructed SR image but also by the estimated blur, this SR feature enhancement encourages the subsequent iterations to intensively learn erroneous features for both better SR and blur kernel estimation.
\item 
In our proposed network, this residual learning is achieved in accordance with a back-projection manner. While the effectiveness of the back projection is demonstrated for non-blind SR with only the SR branch~\cite{haris2018deep}, our proposed network is designed to go along the back-projection manner for blind SR with the blur and SR branches.
This back-projection through both the blur and SR branches improves the estimated blur kernel as well as the SR image.
\end{itemize}


\section{Related work}
\label{section:related_work}

\subsection{Non-kernel SR for a Predefined Degradation}
\label{subsection:nonkernel_sr}

In non-kernel SR, no blur-kernel representation is modeled.
%
Many non-kernel SR methods are proposed with deep networks;
e.g., light-weight networks~\cite{DBLP:conf/eccv/LeeLKH20,DBLP:conf/eccv/LiYLZZYJ20,DBLP:conf/eccv/LuoXZQLF20,DBLP:conf/eccv/XinWJLHG20,DBLP:conf/eccv/LeeDAVK0L20},
SR with reference images~\cite{DBLP:conf/cvpr/ZhangWLQ19,DBLP:conf/eccv/ZhangZDWE020,DBLP:conf/eccv/YanZYZLC20},
edge preservation~\cite{DBLP:journals/tip/YangFYZLGY17,DBLP:conf/cvpr/MaRCCL020},
diverse image reconstruction~\cite{DBLP:conf/cvpr/HeDQ19,DBLP:conf/cvpr/BahatM20,DBLP:conf/eccv/LugmayrDGT20},
data augmentation for SR~\cite{DBLP:conf/cvpr/YooAS20}, attention mechanisms~\cite{DBLP:conf/eccv/ZhangLLWZF18,DBLP:conf/iclr/ZhangLLZF19,DBLP:conf/cvpr/SuganumaLO19,DBLP:conf/cvpr/DaiCZXZ19,DBLP:conf/eccv/NiuWRZYWZCS20}, similarity in feature maps~\cite{johnson2016perceptual}, SR using adversarial training~\cite{ledig2017photo,DBLP:conf/iccv/SajjadiSH17,DBLP:journals/pami/HarisSU21}, iterative/recurrent mechanisms~\cite{haris2018deep,DBLP:journals/pami/HarisSU21,DBLP:journals/tmm/YangMZXYZW19,DBLP:journals/tmm/ZhangWBYZL20}, and implicit image functions~\cite{DBLP:journals/corr/abs-2206-08655}.
Non-kernel SR differs from non-blind SR, where the degradation process is known in inference.
All non-kernel SR methods are non-blind SR because they assume that only a single degradation process (e.g., bicubic downsampling with no blur in Eq.~(\ref{eq:down_model})) is used both in training and inference. 
This assumption inevitably hinders non-kernel SR from getting the best performance for arbitrarily-blurred images.

\subsection{Kernel SR applicable to Any Degradations}
\label{subsection:kernel_sr}

For enabling a single SR model to be applicable to
arbitrarily-degraded images, non-blind SR methods that accept a known blur kernel in inference are developed~\cite{DBLP:conf/iccv/RieglerSRB15,DBLP:conf/cvpr/ZhangZ018,DBLP:conf/cvpr/0008Z019,DBLP:conf/cvpr/XuTTKT20,DBLP:conf/cvpr/ZhangGT20}. 
While they are beneficial as basic techniques and for several applications, this paper focuses on blind kernel SR.

We categorize blind SR methods using blur kernels as follows:
(1) no kernel in networks, 
(2) kernel conditioning, 
and (3) direct image degradation.

\subsubsection{No kernel in networks}
While SR networks in this category have no blur kernel, training or input images are blurred or deblurred by a kernel.
For example, training images degraded by various blurs are used for meta-transfer learning for efficient image-specific SR~\cite{MZSR,DBLP:conf/eccv/ParkYCKK20}.
A more variety of blurs (e.g., compression and sensor noise as well as the general Gaussian noise) improve the SR performance~\cite{BSRGAN2021}.
In SR extended by~\cite{DBLP:conf/eccv/HelouZS20}, training images degraded by various kernels are stochastically masked by different frequency bands for avoiding overfitting to any specific blur kernels.
Various kernels estimated from real images are used
for training in~\cite{DBLP:conf/iccv/ZhouS19}.
A SR network trained with this image set is expected to work well for real images.

Since these methods~\cite{MZSR,DBLP:conf/eccv/ParkYCKK20,DBLP:conf/eccv/HelouZS20,DBLP:conf/iccv/ZhouS19} augment training images for robustness to blurs, this paper focuses more on SR networks with the kernel.
%
While these methods~\cite{MZSR,DBLP:conf/eccv/ParkYCKK20,DBLP:conf/eccv/HelouZS20,DBLP:conf/iccv/ZhouS19} use the kernel to downscale HR training images, correction filter~\cite{hussein2020correction} modifies an input LR image.
The LR image is modified by the kernel estimated from it so that the modified LR image can be super-resolved well by non-kernel SR.
While correction filter works for unknown kernels, it also has no kernel representation in the SR network.

\subsubsection{Kernel conditioning}
For SR robust to various blurs, an SR network is conditioned by a kernel in SRMD~\cite{DBLP:conf/cvpr/ZhangZ018}.
While SRMD is non-blind SR, this kernel conditioning is also useful for blind SR~\cite{DBLP:journals/tog/CornillereDWSS19,DBLP:conf/cvpr/GuLZD19}.
Unlike non-blind SR in which the kernel is given, blind SR must estimate the kernel representation from the LR image.
In~\cite{DBLP:journals/tog/CornillereDWSS19}, the LR image conditioned by the approximated kernel is fed into a network for predicting pixelwise differences between HR and SR images in training.
In inference,
the sum of the differences is minimized to determine the best SR image by optimizing the kernel.
IKC~\cite{DBLP:conf/cvpr/GuLZD19} directly estimates the approximated kernel from the LR image.
The estimated kernel conditions SR features with SFT~\cite{SFT} for feature improvement.
In~\cite{DBLP:conf/cvpr/HuiLW021}, a low-dimensional kernel estimator with reinforcement learning is proposed. 
The estimated low-dimensional kernel is employed by AdaIN~\cite{DBLP:conf/iccv/HuangB17} for controlling the SR process.
DAN~\cite{DBLP:conf/nips/Luo0LWT20} integrates the networks of kernel representation estimation and SR for end-to-end training.
DASR--Wang~\cite{DBLP:conf/cvpr/WangWDX0AG21} improves a degradation kernel representation by contrastive learning~\cite{DBLP:conf/icml/ChenK0H20,DBLP:conf/nips/DosovitskiySRB14,DBLP:conf/cvpr/He0WXG20}.

\subsubsection{Direct image degradation}
The above SR networks conditioned by the approximated kernel have two disadvantages:
(i) the limited accuracy of the approximated kernel and (ii) the indirect degradation process with conditioning.
Rather than unclear conditioning, the model-based representation with Eq. (\ref{eq:down_model}) may improve the representation ability.

%
In KernelGAN~\cite{kernelgan}, the direct degradation expressed by Eq. (\ref{eq:down_model}) is estimated from the LR image based on feature similarity across different scales of the same images.
FKP~\cite{DBLP:conf/cvpr/Liang0GGT21} stabilizes the kernel estimation process in the latent space.
%
MANet~\cite{MANet2021} improves spatially-variant kernel estimation without increasing receptive fields.
%
However, these methods~\cite{kernelgan,DBLP:conf/cvpr/Liang0GGT21,MANet2021} require an additional SR network using the estimated blur kernel.
These kernel estimation and SR networks are not trained in an end-to-end manner.

Such end-to-end training is done in~\cite{DBLP:conf/cvpr/GuoCWCCDXT20,DBLP:conf/cvpr/KimSK21}.
In DRN~\cite{DBLP:conf/cvpr/GuoCWCCDXT20}, an SR image is downscaled to its LR image by the estimated degradation.
In KOALAnet~\cite{DBLP:conf/cvpr/KimSK21}, the raw kernel is estimated by the kernel estimator integrated with the SR network.

However, the aforementioned methods~\cite{kernelgan,DBLP:conf/cvpr/Liang0GGT21,MANet2021,DBLP:conf/cvpr/KimSK21,DBLP:conf/cvpr/GuoCWCCDXT20} employ the estimated degradation just for back-propagation (e.g., with MSE between the input and reconstructed LR images).
In our proposed method, on the other hand, the image degradation is used for iteratively enhancing features for better SR reconstruction using iterative stages.


\section{Preliminary Experiments:
Non-blind SR is improved by various kernels.
}
\label{section:preliminary}

\begin{table}[t]
  \caption{Performance gains of non-kernel SR
    (DBPN~\cite{DBLP:journals/pami/HarisSU21}) by being trained with images degraded by various blur kernels. This method is suffixed by ``-Bl'' (i.e., DBPN-Bl), while DBPN is trained by only images with no blur (i.e., $\sigma \neq 0$).  For comparison, blind kernel SR (IKC~\cite{DBLP:conf/cvpr/GuLZD19}) and its non-blind kernel version (SFTMD~\cite{DBLP:conf/cvpr/GuLZD19}) are shown. The performance is measured by PSNR.  Red and blue scores indicate the best and the second-best, in each column, respectively.
  }
  \label{table:preliminary}
  \begin{center}
    \begin{tabular}{l|ccc}
      \hline
      Method & Set5 & Set14&BSD100\\
      \hline\hline
      DBPN    & 28.82& 25.80 & 26.05 \\
      DBPN-Bl & \blue{29.39}&\blue{26.50} & \blue{26.29} \\
      IKC     & \red{30.92}& \red{27.14}& \red{27.06}\\
      \hline
    \end{tabular}
  \end{center}
\end{table}

We conducted experiments using non-kernel SR with images degraded by various blur kernels.
The aim is to explore how non-kernel SR is improved simply by being trained with these blurred images, as done in blind SR such as~\cite{DBLP:journals/tog/CornillereDWSS19,DBLP:conf/cvpr/GuLZD19}.
While the similar experiments are done in~\cite{bib:xie2021nips}, non-kernel SR and blind SR using kernel are compared with different conditions in~\cite{bib:xie2021nips}. 
That is, in~\cite{bib:xie2021nips}, non-kernel SR is trained with several perceptual losses~\cite{DBLP:conf/eccv/BlauMTMZ18} as well as with a reconstruction MSE loss, but blind SR using kernel is trained with only a reconstruction loss.
Such mixed conditions make it difficult to compare these non-kernel and kernelized methods fairly.
In our experiments, on the other hand, both non-kernel and kernelized methods are trained with only a reconstruction MSE loss for fair comparison.
In addition, we also compare non-kernel SR trained with variously-blurred images with its original one trained only by bicubic downsampling with no blur.
This comparison reveals the effect of the SR training with the variously-blurred images.

Training images are isotropically blurred by Gaussian kernels, each of whose variance is randomly selected from $[ 0.2, 4.0 ]$. 
Each test image is also isotropically blurred by the Gaussian kernel $\sigma \in \{ 1.3, 2.6 \}$.
The results are shown in Table~\ref{table:preliminary}.
We can see that DBPN-Bl (trained by various blur kernels) is much better than the original DBPN trained only by bicubic downsampling with no blur.
DBPN-Bl is superior or comparable to IKC~\cite{DBLP:conf/cvpr/GuLZD19}, which is one of the blind SR methods.

These results prod us into (i) revisiting a high-performance non-blind SR method (e.g., DBPN) and (ii) extending it to blind SR with blur kernels.


\section{Proposed Methods}
\label{section:method}

For organically integrating blind kernel estimation into any high-performance non-blind SR method, these blind and non-blind SR networks should have similar structures. 
Among potential combinations, we pay attention to iterative networks because of their abilities for augmenting and gradually revising features~\cite{haris2018deep,DBLP:journals/pami/HarisSU21,DBLP:conf/cvpr/GuLZD19,DBLP:conf/cvpr/GuoCWCCDXT20}.
Based on this motivation, this paper proposes the following two methods.

In the first method, kernel conditioning is integrated with an iterative end-to-end SR network, consisting of up- and down-projection modules~\cite{haris2018deep}, as shown in Fig.~\ref{fig:compare_sr}~\subref{fig:kcbpn}. 
For stabilizing both kernel estimation and conditioning, low-dimensional approximated kernels are employed, as proposed in~\cite{DBLP:journals/tog/CornillereDWSS19,DBLP:conf/cvpr/GuLZD19}.
Different from IKC~\cite{DBLP:conf/cvpr/GuLZD19} (Fig.~\ref{fig:compare_sr}~\subref{fig:ikc_sftmd}) where the iterative network is also used, the SR branch in our proposed network is based on iterative up-and-down samplings for SR feature enhancement with back projections~\cite{haris2018deep,DBLP:journals/pami/HarisSU21}.
Furthermore, IKC consists of three separate networks, which are not trained in an end-to-end manner.
We call this network the {\em Kernel-Conditioned Back-Projection Network} (KCBPN).

Since KCBPN is a simple integration of IKC and DBPN, KCBPN lacks several important components of kernelized SR introduced in Sec.~\ref{section:related_work}.
First, its kernel representation is an approximated low-dimensional vector that may result in limited SR accuracy.
Second, the approximated kernel makes it impossible to directly employ the image degradation expressed by Eq. (\ref{eq:down_model}).
These problems are resolved in our second method (Fig.~\ref{fig:compare_sr}~\subref{fig:kbpn}), which is called the {\em Kernelized Back-Projection Network} (KBPN).
Unlike previous blind SR networks with the image degradation in which blurring and downsampling are integrated into one process~\cite{kernelgan,DBLP:conf/cvpr/Liang0GGT21,MANet2021,DBLP:conf/cvpr/KimSK21,DBLP:conf/cvpr/GuoCWCCDXT20}, KBPN models Eq. (\ref{eq:down_model}) by separating blurring and downsampling sub-processes.
This separate modeling with the simpler sub-processes allows us to simplify blind SR learning. 
%
More importantly, we propose to employ a residual between the reconstructed LR and its input LR images.
While this residual is used for only back-propagation in previous blind SR methods, our method propagates it forward for learning erroneous features in the residual through the iterative mechanism.

In both KCBPN and KBPN, each iterative stage consists of upsampling, downsampling, blur estimation, and image degradation processes.
SR features extracted in all iterative stages (denoted by $F^{SR}_{1,\cdots,T}$ where $T$ denotes the number of stages) are finally concatenated channelwise and fed into the reconstruction layer with $3 \times 3$ convolution, as depicted by red dotted arrows in Fig.~\ref{fig:compare_sr}~\subref{fig:kcbpn} and \subref{fig:kbpn}, for the final SR image (denoted by $I^{SR}$).

\subsection{Kernel Conditioned Back-Projection Networks}
\label{subsection:kcbpn}

\begin{figure}[t]
  \begin{center}
    \includegraphics[width=1.\columnwidth]{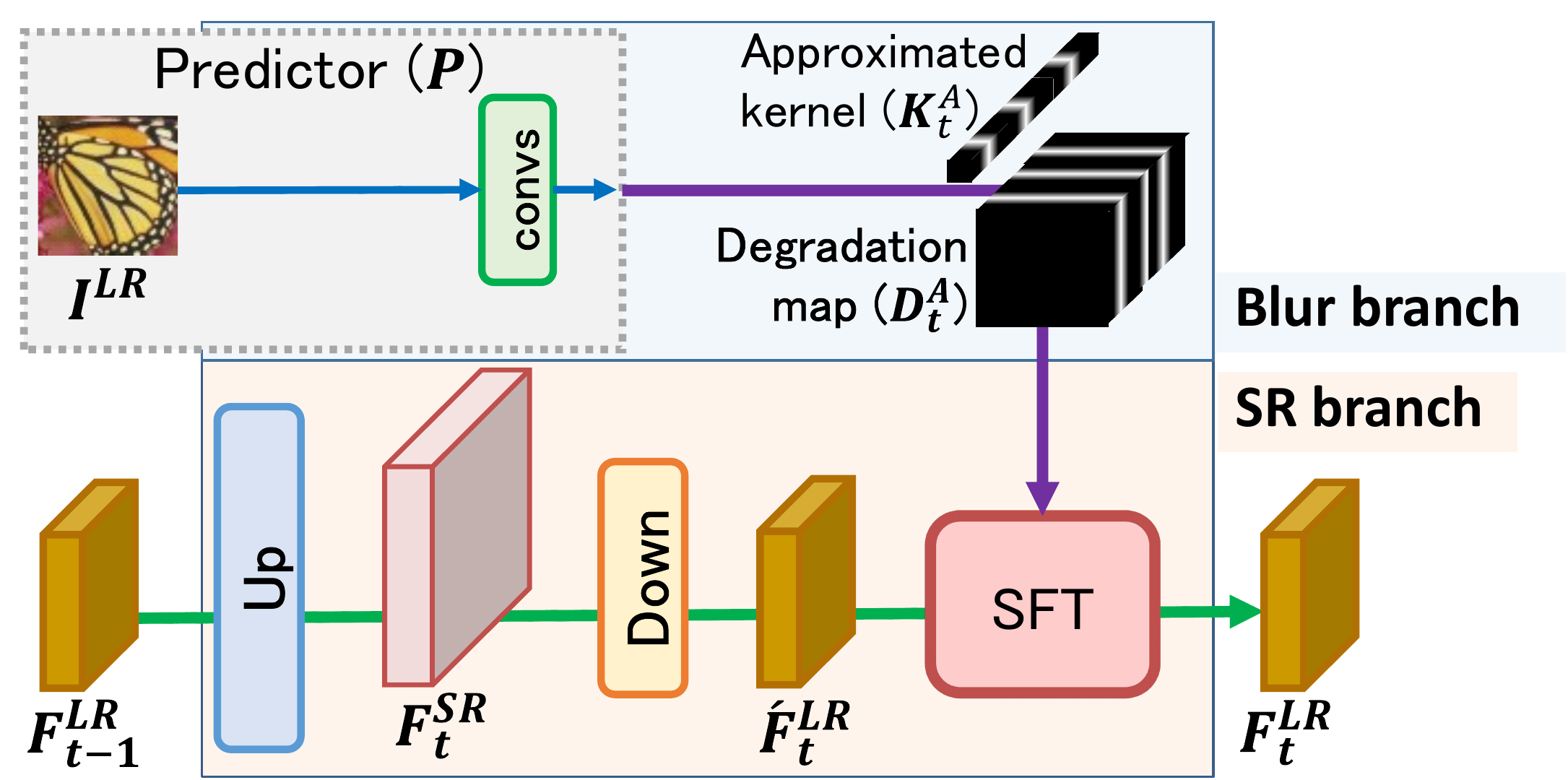}
    \caption{$t$-th stage in KCBPN.}
    \label{fig:kcbpn_stage}
  \end{center}
\vspace*{5mm}
  \begin{center}
    \includegraphics[width=1.\columnwidth]{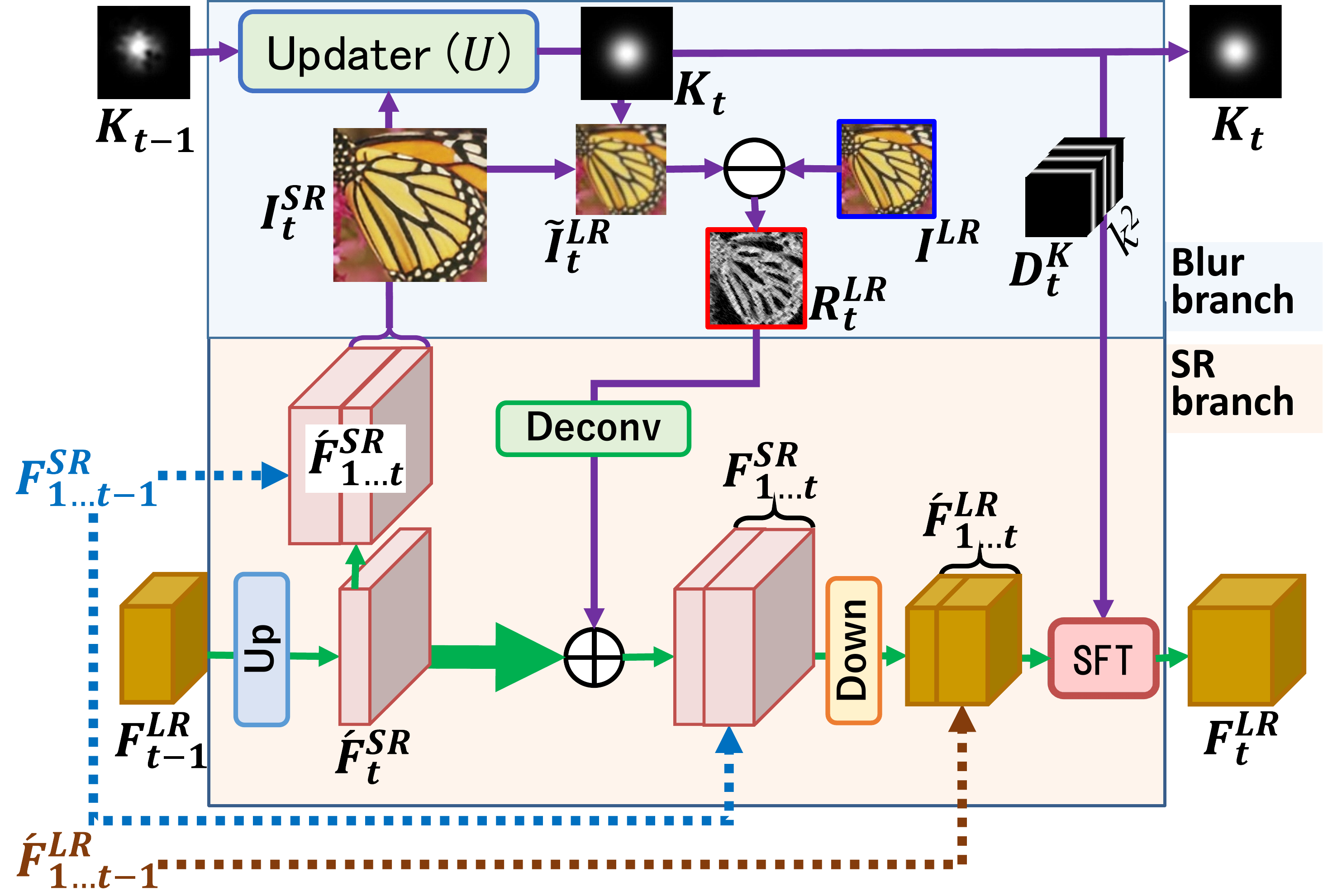}
    \caption{$t$-th stage in KBPN.}
    \label{fig:kbpn_stage_sr_kbpn2}
  \end{center}
\end{figure}

The overall structure and each stage in KCBPN are illustrated in Figs.~\ref{fig:compare_sr}~\subref{fig:kcbpn} and \ref{fig:kcbpn_stage}, respectively.
``LR features extracted in the $(t-1)$-th stage (denoted by $F^{LR}_{t-1}$)'' and ``the degradation map (denoted by $D^{A}_{t}$) extracted from an input LR image'' are fed into the $t$-th stage.
Its output is LR features (denoted by $F^{LR}_{t}$).
$D^{A}_{t}$ is the stretch of the low-dimensional vector of the kernel given by the kernel predictor, $P$.

In the SR branch, $F^{LR}_{t-1}$ is up- and down-sampled for feature augmentation by extracting SR features, $F^{SR}_{t}$, and LR features, $\acute{F}^{LR}_{t}$, as proposed in~\cite{haris2018deep}.
$\acute{F}^{LR}_{t}$ is conditioned with $D^{A}_{t}$ by SFT~\cite{SFT} as follows:
\begin{eqnarray}
    \gamma & = & S({\rm convs1}({\rm concat}(\acute{F}^{LR}_{t}, D^{A}_{t}))) \nonumber \\
    \beta & = & {\rm convs2}({\rm concat}(\acute{F}^{LR}_{t}, D^{A}_{t})) \nonumber \\
    F^{LR}_{t} &=& \acute{F}^{LR}_{t} \odot \gamma \oplus \beta,
    \label{eq:sft}
\end{eqnarray}
where ${\rm convs1}$ and ${\rm convs2}$ are two-layer CNNs.
Their kernel size is $3 \times 3$.
Each conv layer is followed by LeakyReLU ($\alpha=0.1$).
$S(\cdot)$ denotes the sigmoid function.

\subsection{Kernelized Back-Projection Networks}
\label{subsection:kbpn}

The overall structure and each stage of KBPN are shown in Fig.~\ref{fig:compare_sr} \subref{fig:kbpn} and Fig.~\ref{fig:kbpn_stage_sr_kbpn2}, respectively.
As with KCBPN, (i) the SR branch has up- and down-projection modules, and (ii) the kernel is fed into the SR branch from the blur branch.
On the other hand, different from similar methods such as IKC and KCBPN, an SR image reconstructed in each stage is blurred by an estimated raw kernel $K_{t}$ ($k\times k$ pixels) and downsampled by Eq.~(\ref{eq:down_model}).
KBPN consists of the following four steps.

In Step 1 of the $t$-th stage, the up-projection module extracts SR
features from $F^{LR}_{t-1}$ (i.e., LR features given by the $(t-1)$-th stage).
The upscaled features (denoted by $\acute{F}^{SR}_{t}$) are concatenated channelwise with $F^{SR}_{1,\cdots,t-1}$ given by the previous stages.
%
The concatenated features are denoted by $\acute{F}^{SR}_{1,\cdots,t}$.

In Step 2, $\acute{F}^{SR}_{1,\cdots,t}$ is used for reconstructing $I^{SR}_{t}$.
As with the final reconstruction layer, this SR reconstruction is done by a $3 \times 3$ convolution layer.
$I^{SR}_{t}$ and $K_{t-1}$ are fed into the blur updater,
$U$, for updating $K_{t-1}$ to $K_{t}$.  $I^{SR}_{t}$ is convolved by
$K_{t}$ for estimating $\tilde{I}^{LR}_{t}$.

Step 3 updates features by the residual between $\tilde{I}^{LR}_{t}$
and the input LR image ($I^{LR}$).
This residual, $R^{LR}_{t} = \tilde{I}^{LR}_{t} - I^{LR}$, represents how well $I^{SR}_{t}$ and $K_{t}$ are reconstructed.
We propose to employ $R^{LR}_{t}$ for enhancing SR features in the following iterations (i.e., $F^{SR}_{\tau}$ where $\tau = \{ t+1, t+2, \cdots, T \}$).
For this {\bf SR feature enhancement}, $R^{LR}_{t}$ is provided from the blur branch to the SR branch so that $\acute{F}^{SR}_{t}$ is enhanced.
Let this enhanced SR feature be $F^{SR}_{t}$.
We propose to design this SR feature enhancement as residual learning in which $\acute{F}^{SR}_{t}$ is given through the identical mapping, indicated by the thick green arrow in Fig.~\ref{fig:kbpn_stage_sr_kbpn2}.
To this end, $R^{LR}_{t}$ is upsampled by two conv layers and a deconv layer (indicated by ``Deconv'' in Fig.~\ref{fig:kbpn_stage_sr_kbpn2}) for adding it to $\acute{F}^{SR}_{t}$ elementwise.
This residual feature feedback is done by elementwise-add in accordance with the original back-projection~\cite{irani1991improving} and its network implementation~\cite{haris2018deep}, while they~\cite{irani1991improving,haris2018deep} are designed with only the SR branch without the blur branch.
For this elementwise-add operation, the channel size of $R^{LR}_{t}$ is changed in this Deconv so that its channel size is equal to that of $\acute{F}^{SR}_{t}$.
The kernel sizes of the first and second conv layers are $3 \times 3$ and $1 \times 1$, respectively.
The enhanced SR features ($F^{SR}_{t}$) is then concatenated channelwise with $F^{SR}_{1,\cdots,t-1}$ given by the previous stages.
This feature concatenation is inspired by dense connections in D-DBPN~\cite{haris2018deep} to alleviate the vanishing gradient problem and encourage feature reuse.
The elementwise-added features (denoted by $F^{SR}_{1,\cdots,t}$) are fed into the down-projection module for getting the LR feature, $\acute{F}^{LR}_{1,\cdots,t}$.

In Step 4, the updated kernel, $K_{t}$, is vectorized to a
$k^{2}$-dimensional vector and stretched to the degradation maps
(denoted by $D^{K}_{t}$), as proposed
in~\cite{DBLP:conf/cvpr/ZhangZ018}.  $D^{K}_{t}$ is fed into SFT as
the condition for updating $\acute{F}^{LR}_{1,\cdots,t}$ to $F^{LR}_{t}$ by
Eq. (\ref{eq:sft}).  Finally, $F^{LR}_{t}$ and $K_{t}$ are provided to
the $(t+1)$-th stage.

The technical novelties in KBPN compared to the most related blind SR
methods are summarized as follows:
\begin{itemize}
\item In SRMD~\cite{DBLP:conf/cvpr/ZhangZ018},
  AMNet~\cite{DBLP:conf/cvpr/HuiLW021},
  DAN~\cite{DBLP:conf/nips/Luo0LWT20},
  DASR--Wang~\cite{DBLP:conf/cvpr/WangWDX0AG21}, and
  IKC~\cite{DBLP:conf/cvpr/GuLZD19}, a low-dimensional kernel
  representation is used only for conditioning LR features, while KBPN
  also conditions the LR features by the estimated raw kernel.
\item 
  A raw kernel estimated in DRN~\cite{DBLP:conf/cvpr/GuoCWCCDXT20} and KOALAnet\cite{DBLP:conf/cvpr/KimSK21} is used only for downscaling the reconstructed SR image.
  Then, a residual between the downscaled image and the input LR image is computed for back-propagation.
  In addition to this back-propagation, KBPN uses this residual (i.e., $R^{LR}_{t}$) also for enhancing SR features.
  Since pixels each of which has a higher value in $R^{LR}_{t}$ can be regarded as erroneous pixels in $\tilde{I}^{LR}_{t}$, conditioning by $R^{LR}_{t}$ encourages the next stage to explicitly learn features in order to focus on reconstructing these erroneous pixels.
  Integrating this feature enhancement with a back-projection manner~\cite{irani1991improving,haris2018deep} between the blur and SR branches is our original scheme.
\item Iterations in IKC~\cite{DBLP:conf/cvpr/GuLZD19} gradually get the kernel and SR image closer to their ground-truths.
  On the other hand, KBPN iterates the stages for feature enhancement (i.e., for producing different features) and utilizes the features extracted in all the stages for SR reconstruction.
  Furthermore, IKC is not trained in an end-to-end manner.
\end{itemize}

\subsection{Loss Functions}
\label{subsection:loss}

In both KCBPN and KBPN, the following MSE loss is used for the SR
image:
\begin{footnotesize}
\begin{equation}
    \mathcal{L}_{SR} = \frac{1}{CHW} \sum_c^{C} \sum_{h}^{H} \sum_{w}^{W} \left( I^{SR}(c,h,w) - I^{HR}(c,h,w) \right)^{2},
    \label{eq:sr_mse}
\end{equation}
\end{footnotesize}
\hspace*{-1.5mm}where $C$, $H$, and $W$ denote the channel size, height, and width of the HR and SR images, respectively.

The kernels estimated in the last $T$-th stage are considered to be the final estimations so that $K^{A} = K^{A}_{T}$ and $K = K_{T}$.
These kernels are used for the kernel loss, denoted by $\mathcal{L}_{KC}$ and $\mathcal{L}_{K}$ in KCBPN and KBPN, respectively:
\begin{eqnarray}
    \mathcal{L}_{KC} &=& \frac{1}{a} \sum_{i}^{a} | K^{A}(i) - \bar{K}^{A}(i) |,\\
    \label{eq:kc_loss}
    \mathcal{L}_{K} &=& \frac{1}{k^{2}} \sum_{j}^{k} \sum_{i}^{k} | K(j,i) - \bar{K}(j,i) |,
    \label{eq:k_loss}
\end{eqnarray}
where $\bar{K}^{A}$ and $\bar{K}$ denote the ground-truth of $K^{A}$ in KCBPN and $K$ in KBPN, respectively.
The dimension of $K^{A}$ is $a$.

In addition, we propose to evaluate the LR image degraded from the output SR image $I^{SR}$ in KBPN.
This LR image is $\tilde{I}^{LR}_{T} = (I^{SR} \ast K)\downarrow_{s}$.
The loss defined by the MSE between $\tilde{I}^{LR}$ and $I^{LR}$ is called the {\bf LR loss}:

\begin{footnotesize}
\begin{equation}
    \mathcal{L}_{LR} = \frac{s^{2}}{CHW} \sum_{c}^{C} \sum_{h}^{H/s} \sum_{w}^{W/s} \left( \tilde{I}^{LR}_{T}(c,h,w) - I^{LR}(c,h,w) \right)^{2}
    \label{eq:lr_mse}
\end{equation}
\end{footnotesize}

\noindent
While the LR loss is used also in~\cite{DBLP:conf/cvpr/YuanLZZDL18,DBLP:conf/cvpr/GuoCWCCDXT20}, these methods reconstruct the LR image from the SR image by non-linear downsampling neglecting the image degradation equation~\ref{eq:down_model}.
In our method, on the other hand, the LR image is reconstructed by~\ref{eq:down_model}, which allows us to separate the whole degradation to the blurring and downsampling sub-processes for easier optimizations of these sub-processes.

In both KCBPN and KBPN, the weights of the SR loss ($\mathcal{L}_{SR}$) and the kernel loss ($\mathcal{L}_{KC}$ in KCBPN and $\mathcal{L}_{K}$ in KBPN) are 1 and 5, respectively.
In KBPN, the weight of the LR loss, $\mathcal{L}_{LR}$, is 0.1.

\begin{figure}[t]
  \begin{center}
    \includegraphics[width=1.\columnwidth]{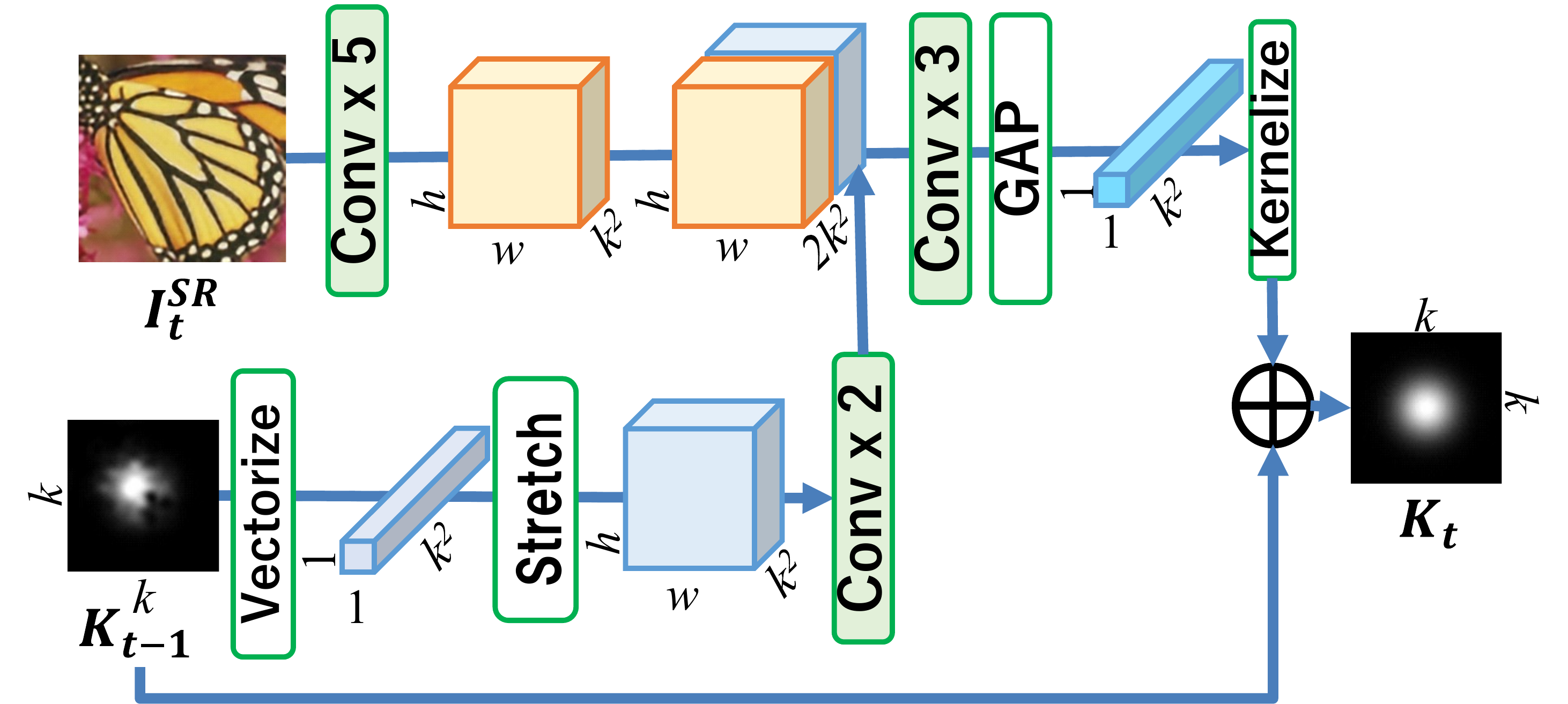}
    \caption{Blur updater in KBPN.}
    \label{fig:updater}
  \end{center}
\end{figure}

\begin{figure}[t]
  \begin{center}
    \includegraphics[width=1.\columnwidth]{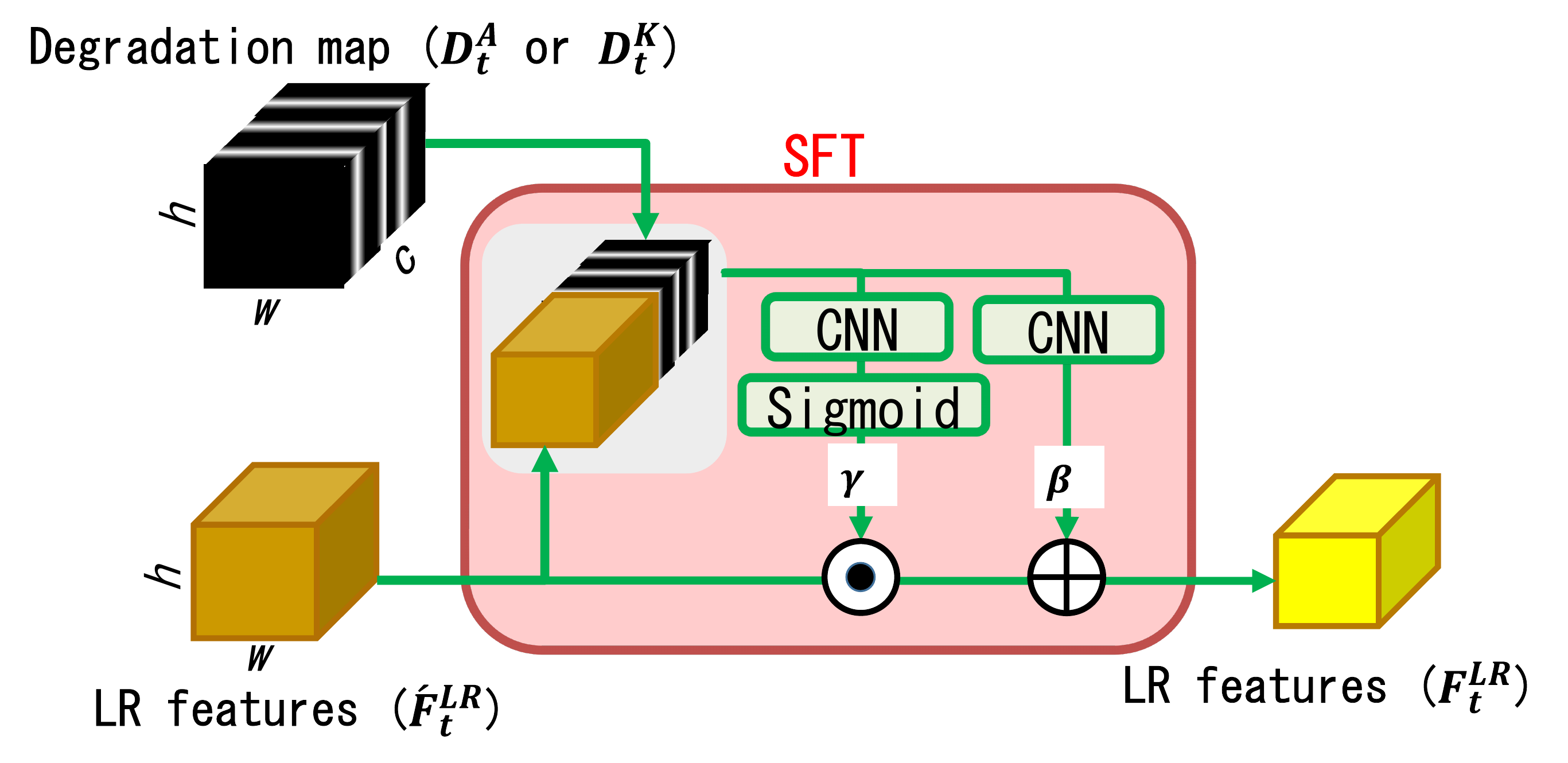}
    \caption{Detail of SFT used in KCBPN and KBPN.}
    \label{fig:sft_kernel_condition}
  \end{center}
\end{figure}


\begin{table*}[t]
  \caption{Comparison with blind SR methods. Both training and test images are blurred by isotropic blur kernels.
  In each column, \red{red}, \blue{blue}, and \green{green} denote the best, second-best, and third-best scores, respectively.}
  \label{table:ex_comparison}
  \begin{center}
  \begin{footnotesize}
    \begin{tabular}{l||c|cccccc||c|cccccc}\hline
      \multirow{2}{*}{Method} &
      \multirow{2}{*}{\hspace*{-1mm}$\sigma$\hspace*{-1mm}}& \multicolumn{2}{|c}{Set5}&\multicolumn{2}{c}{Set14}&\multicolumn{2}{c||}{BSD100}&
      \multirow{2}{*}{\hspace*{-1mm}$\sigma$\hspace*{-1mm}}& \multicolumn{2}{|c}{Set5}&\multicolumn{2}{c}{Set14}&\multicolumn{2}{c}{BSD100}\\
      &&PSNR&SSIM  &PSNR&SSIM  &PSNR&SSIM && PSNR&SSIM  &PSNR&SSIM  &PSNR&SSIM\\ \hline\hline

      DBPN~\cite{DBLP:journals/pami/HarisSU21} &\multirow{6}{*}{\hspace*{-1mm}0.2\hspace*{-1mm}} & 23.44 & 0.714 & 19.91 & 0.584 & 20.52 & 0.570 & \multirow{6}{*}{\hspace*{-1mm}2.6\hspace*{-1mm}} & 26.80 & 0.772 & 24.90 & 0.660 & 25.17 & 0.630\\
      DBPN-Bl                                 & & \green{26.45} & \green{0.809} & \green{22.15} & \green{0.669} & \green{23.02} & \green{0.655} &  & 28.05 & 0.804 & 25.87 & 0.698 & 25.60 & 0.648\\
      IKC~\cite{DBLP:conf/cvpr/GuLZD19}       & & 24.33 & 0.757 & 19.78 & 0.600 & 20.66 & 0.595&                 & \blue{30.93} & \blue{0.870} & \blue{27.35} & \blue{0.742} & \red{27.19} & \red{0.715}  \\
      DAN~\cite{DBLP:conf/nips/Luo0LWT20}     & & 25.44 & 0.777 & 20.92 & 0.589 & 21.76 & 0.609 &                 & \green{29.56} & \green{0.848} & \green{26.31} & \green{0.726} & \green{26.73} & \green{0.701} \\
      KCBPN (stg7)                                   & & \blue{30.18} & \blue{0.873} & \blue{27.43} & \blue{0.768} & \blue{26.84} & \blue{0.723} & & 28.27 & 0.803 & 26.03 & 0.698 & 25.69 & 0.644\\
      KBPN (stg7)                                    & & \red{31.83} & \red{0.890} & \red{28.21} & \red{0.777} & \red{27.29} & \red{0.728} & & \red{31.21} & \red{0.871} & \red{28.07} & \red{0.756} & \blue{27.14} & \blue{0.705}\\
      \hline
      DBPN~\cite{DBLP:journals/pami/HarisSU21} &\multirow{6}{*}{\hspace*{-1mm}1.3\hspace*{-1mm}} & \green{30.83} & \blue{0.879} & 26.70 & \green{0.752} & 26.93 & \blue{0.729} & \multirow{6}{*}{\hspace*{-1mm}4.0\hspace*{-1mm}} &24.23 & 0.671 & 23.05 & 0.576 & 23.67 & 0.555\\
      DBPN-Bl                                 & & 30.72 & 0.875 & \green{27.13} & 0.751 & \green{26.98} & 0.725& & 24.53 & 0.678 & 23.33 & 0.585 & \green{23.82} & 0.559\\
      IKC~\cite{DBLP:conf/cvpr/GuLZD19}       & & \blue{30.90} & \green{0.878} & 26.92 & 0.751 & 26.92 & 0.726 & & \blue{25.75} & \red{0.740} & \blue{24.23} & \blue{0.622} & \blue{24.54} & \red{0.601} \\
      DAN~\cite{DBLP:conf/nips/Luo0LWT20}     & & \blue{30.90} & \blue{0.879} & 26.91 & 0.751 & 26.96 & \green{0.727}&                 & \green{25.13} & \green{0.705} & \green{23.83} & \green{0.609} & 24.36 & 0.585 \\
      KCBPN (stg7)                                   & & 30.78 & 0.876 & \blue{27.80} & \blue{0.767} & \blue{27.15} & 0.720 & & 25.11 & 0.696 & 23.61 & 0.593 & 24.06 & \green{0.565}\\
      KBPN (stg7)                                    & & \red{32.33} & \red{0.895} & \red{28.60} & \red{0.783} & \red{27.61} & \red{0.733} & & \red{26.15} & \blue{0.736} & \red{24.67} & \red{0.635} & \red{24.76} & \blue{0.594}\\
      \hline

    \end{tabular}
  \end{footnotesize}
  \end{center}
\end{table*}

\begin{table*}[t]
  \caption{Comparison with blind image-specific SR methods.}
  \label{table:ex_comparison_fully_blind}
  \begin{center}
  \begin{footnotesize}
  \begin{tabular}{l||c|cccc||c|cccc}\hline
      \multirow{2}{*}{Method} & \multirow{2}{*}{$\sigma$} &\multicolumn{2}{c}{Set14~\cite{set14}}&\multicolumn{2}{c||}{BSD100~\cite{BSDS100}}& \multirow{2}{*}{$\sigma$} &\multicolumn{2}{c}{Set14~\cite{set14}}&\multicolumn{2}{c}{BSD100~\cite{BSDS100}}\\
      & &PSNR   &SSIM   &PSNR   &SSIM&  &PSNR   &SSIM   &PSNR   &SSIM \\ \hline\hline
      (a) ZSSR~\cite{ZSSR} + KernelGAN~\cite{kernelgan} (Fully-blind)     &\multirow{4}{*}{$\frac{3.5}{\sqrt{2}}$} & 24.42 & 0.673 & 24.36 & 0.647&\multirow{4}{*}{$\frac{4.5}{\sqrt{2}}$} & 25.17 & 0.669 & 25.24 & 0.652\\
      (b) DBPN + Correction~\cite{hussein2020correction} (Fully-blind)    &  & \blue{28.18} & \red{0.764} & \blue{27.10} & \red{0.722}&         & \blue{25.54} & \blue{0.699} & \blue{25.49} & \red{0.671} \\
      (c) KCBPN (stg7)                                                           &  & 26.31 & \blue{0.709} & 25.85 & 0.652& & 24.90 & 0.649 & 24.96 & 0.608\\
      (d) KBPN (stg7)                                                             &  & \red{28.26} & \red{0.764} & \red{27.29} & \blue{0.713} & & \red{26.76} & \red{0.710} & \red{26.23} & \blue{0.660}\\
      \hline
    \end{tabular}
  \end{footnotesize}
  \end{center}
\end{table*}

\section{Experimental Results}
\label{section:experiments}

\subsection{Detailed Implementation and Datasets}
\label{subsection:setting}

As with~\cite{DBLP:conf/cvpr/ZhangWLQ19}, in both KCBPN and KBPN, VGG-16~\cite{DBLP:journals/corr/SimonyanZ14a} is employed for feature extraction from an input LR image.
More specifically, the first four convolution layers of VGG-16 without the pooling layers extract initial LR features, $F^{LR}_{0}$, from $I^{LR}$.
The kernel sizes of the first and second conv layers are $3 \times 3$ and $1 \times 1$, respectively.
Up- and down-projection modules are designed in the same way as DBPN~\cite{haris2018deep}.
The reconstruction layer is a $3 \times 3$ conv layer with no activation.
The sizes of initial LR and SR features (i.e., $F^{LR}_{1}$ and $F^{SR}_{1}$) are $(c, w, h) = (64, \frac{W}{s}, \frac{H}{s})$ and $(c, w, h) = (64, W, H)$, respectively.
The blur-kernel size $k$ is 21.
The kernel predictor $P$ consisting of four convolutional layers and the last global average pooling layer in KCBPN and KBPN is identical to the one used in IKC~\cite{DBLP:conf/cvpr/GuLZD19}.
Given the blur kernel with $k \times k$ pixels, the channel size of the output vector (denoted by $c$) is $c = 9$ and $c = k^{2} = 441$ in KCBPN and KBPN, respectively.
In KBPN, a $k^{2}$-dimensional vector estimated by this blur predictor is reshaped to a $k \times k$ matrix.
The architecture of the blur updater $U$ in KBPN is shown in Fig.~\ref{fig:updater}.

The detailed architectures of SFT~\cite{SFT} is illustrated in Fig.~\ref{fig:sft_kernel_condition}.
For SFT, the degradation map (i.e., $D^{A}_{t}$ in KCBPN and $D^{K}_{t}$ in KBPN) and the LR feature (i.e., $\acute{F}^{LR}_{t}$) are concatenated channelwise.
This concatenated feature is fed into two convolution layers to get $\gamma$ and $\beta$ in parallel.
The kernel size of each convolution layer is $3 \times 3$.
Each convolution layer is followed by LeakyReLU ($\alpha = 0.1$).
With $\gamma$ and $\beta$, the LR feature is updated to $F^{LR}_{t}$ using Eq. (\ref{eq:sft}).

We trained and evaluated two types of SR models independently.
The first and second models were trained with images blurred by isotropic and anisotropic Gaussian kernels, respectively.
The numbers of the iterative stages are 7 and 4 for isotropic and anisotropic Gaussian blur kernels, respectively, in both KCBPN  and KBPN.
In addition, KBPN is evaluated also with 3 and 7 stages for anisotropic Gaussian blur kernels.
All experiments were conducted with a scale factor of 4.

\noindent
{\bf Training:}
3,450 images in DIV2K~\cite{DIV2K} and Flickr2K~\cite{Flickr2K} were used.
All images were horizontally and vertically flipped for augmentation.
The variance of the Gaussian blur kernel is randomly selected from $[ 0.2, 4.0 ]$.
%
%
Adam~\cite{DBLP:journals/corr/KingmaB14} was used as an optimizer.
The learning rate was initially $10^{-4}$ and reduced to $10^{-5}$.
The batch size was 8.

\begin{figure*}
  \begin{center}
    \begin{tabular}{c}
      \begin{minipage}{1\linewidth}
       \begin{minipage}{0.116\linewidth}
         \begin{center}
          \includegraphics[width=1\linewidth]{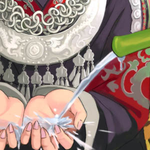}
~
        \end{center}
       \end{minipage}
       \begin{minipage}{0.116\linewidth}
        \begin{center}
          \includegraphics[width=1\linewidth]{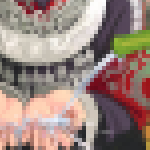}
~
        \end{center}
       \end{minipage}
       \begin{minipage}{0.116\linewidth}
        \begin{center}
          \includegraphics[width=1\linewidth]{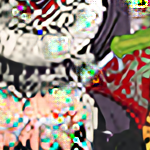}
(15.41)
        \end{center}
      \end{minipage}
       \begin{minipage}{0.116\linewidth}
        \begin{center}
          \includegraphics[width=1\linewidth]{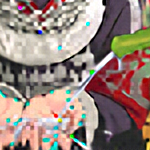}
(17.52)
        \end{center}
       \end{minipage}
       \begin{minipage}{0.116\linewidth}
        \begin{center}
          \includegraphics[width=1\linewidth]{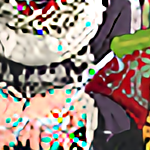}
(14.73)
        \end{center}
       \end{minipage}
       \begin{minipage}{0.116\linewidth}
        \begin{center}
          \includegraphics[width=1\linewidth]{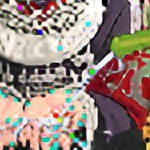}
(16.41)
        \end{center} 
       \end{minipage}
       \begin{minipage}{0.116\linewidth}
        \begin{center}
          \includegraphics[width=1\linewidth]{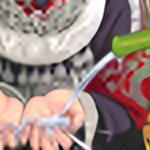}
\red{(20.47)}
        \end{center} 
       \end{minipage}
       \begin{minipage}{0.116\linewidth}
        \begin{center}
          \includegraphics[width=1\linewidth]{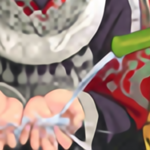}
\blue{(20.45)}
        \end{center} 
       \end{minipage}
     \end{minipage}\\
     \begin{minipage}{1\linewidth}
       \begin{minipage}{0.116\linewidth}
         \begin{center}
          \includegraphics[width=1\linewidth]{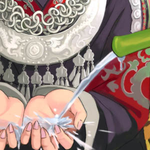}
~
        \end{center}
       \end{minipage}
       \begin{minipage}{0.116\linewidth}
        \begin{center}
          \includegraphics[width=1\linewidth]{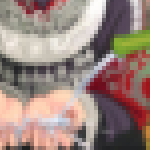}
~
        \end{center}
       \end{minipage}
       \begin{minipage}{0.116\linewidth}
        \begin{center}
          \includegraphics[width=1\linewidth]{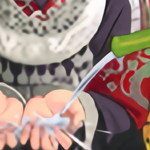}
(20.21)
        \end{center}
      \end{minipage}
       \begin{minipage}{0.116\linewidth}
        \begin{center}
          \includegraphics[width=1\linewidth]{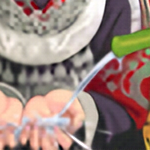}
(20.59)
        \end{center}
       \end{minipage}
       \begin{minipage}{0.116\linewidth}
        \begin{center}
          \includegraphics[width=1\linewidth]{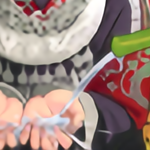}
(20.43)
        \end{center}
       \end{minipage}
       \begin{minipage}{0.116\linewidth}
        \begin{center}
          \includegraphics[width=1\linewidth]{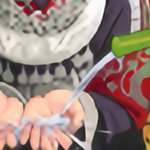}
(20.43)
        \end{center} 
       \end{minipage}
       \begin{minipage}{0.116\linewidth}
        \begin{center}
          \includegraphics[width=1\linewidth]{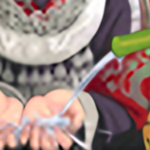}
\red{(20.71)}
        \end{center} 
       \end{minipage}
       \begin{minipage}{0.116\linewidth}
        \begin{center}
          \includegraphics[width=1\linewidth]{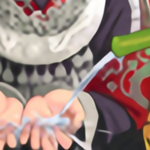}
\blue{(20.68)}
        \end{center} 
       \end{minipage}
     \end{minipage}\\
     \begin{minipage}{1\linewidth}
       \begin{minipage}{0.116\linewidth}
         \begin{center}
          \includegraphics[width=1\linewidth]{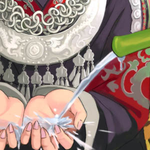}
~
        \end{center}
       \end{minipage}
       \begin{minipage}{0.116\linewidth}
        \begin{center}
          \includegraphics[width=1\linewidth]{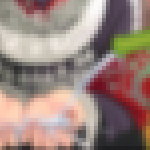}
~
        \end{center}
       \end{minipage}
       \begin{minipage}{0.116\linewidth}
        \begin{center}
          \includegraphics[width=1\linewidth]{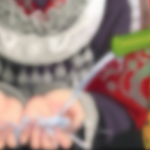}
(19.63)
        \end{center}
      \end{minipage}
       \begin{minipage}{0.116\linewidth}
        \begin{center}
          \includegraphics[width=1\linewidth]{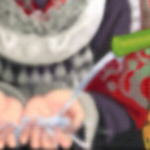}
(20.15)
        \end{center}
       \end{minipage}
       \begin{minipage}{0.116\linewidth}
        \begin{center}
          \includegraphics[width=1\linewidth]{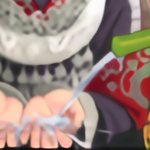}
(20.42)
        \end{center}
       \end{minipage}
       \begin{minipage}{0.116\linewidth}
        \begin{center}
          \includegraphics[width=1\linewidth]{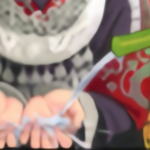}
\blue{(20.48)}
        \end{center} 
       \end{minipage}
       \begin{minipage}{0.116\linewidth}
        \begin{center}
          \includegraphics[width=1\linewidth]{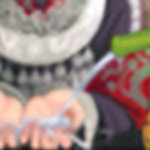}
(20.06)
        \end{center} 
       \end{minipage}
       \begin{minipage}{0.116\linewidth}
        \begin{center}
          \includegraphics[width=1\linewidth]{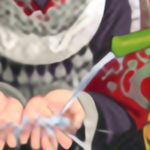}
\red{(20.77)}
        \end{center} 
       \end{minipage}
     \end{minipage}\\
     \begin{minipage}{1\linewidth}
       \begin{minipage}{0.116\linewidth}
         \begin{center}
          \includegraphics[width=1\linewidth]{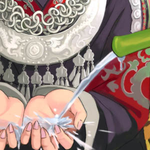}
~\\\subcaption{HR}
          \label{fig:sigma_40_set14_005_hr}
        \end{center}
       \end{minipage}
       \begin{minipage}{0.116\linewidth}
        \begin{center}
          \includegraphics[width=1\linewidth]{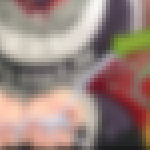}
~\\ \subcaption{LR}
          \label{fig:sigma_40_set14_005_input}
        \end{center}
       \end{minipage}
       \begin{minipage}{0.116\linewidth}
        \begin{center}
          \includegraphics[width=1\linewidth]{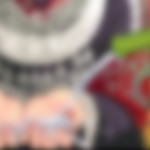}
(18.33)\\ \subcaption{DBPN}
          \label{fig:sigma_40_set14_005_dbpn}
        \end{center}
      \end{minipage}
       \begin{minipage}{0.116\linewidth}
        \begin{center}
          \includegraphics[width=1\linewidth]{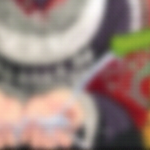}
(18.53)\\ \subcaption{DBPN-Bl}
          \label{fig:sigma_40_set14_005_dbpn_bl}
        \end{center}
       \end{minipage}
       \begin{minipage}{0.116\linewidth}
        \begin{center}
          \includegraphics[width=1\linewidth]{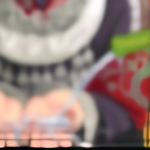}
\blue{(18.88)}\\ \subcaption{IKC}
          \label{fig:sigma_40_set14_005_ikc}
        \end{center}
       \end{minipage}
       \begin{minipage}{0.116\linewidth}
        \begin{center}
          \includegraphics[width=1\linewidth]{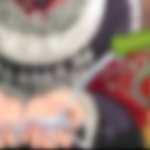}
(18.51)\\ \subcaption{DAN}
          \label{fig:sigma_40_set14_005_dan}
        \end{center} 
       \end{minipage}
       \begin{minipage}{0.116\linewidth}
        \begin{center}
          \includegraphics[width=1\linewidth]{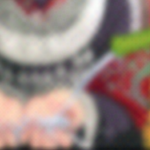}
(18.65)\\ \subcaption{KCBPN}
          \label{fig:sigma_40_set14_005_kcbpn}
        \end{center} 
       \end{minipage}
       \begin{minipage}{0.116\linewidth}
        \begin{center}
          \includegraphics[width=1\linewidth]{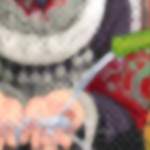}
\red{(19.75)}\\ \subcaption{KBPN}
          \label{fig:sigma_40_set14_005_kbpn}
        \end{center} 
       \end{minipage}
     \end{minipage}\\
     \\
     \\
     \begin{minipage}{1\linewidth}
       \begin{minipage}{0.116\linewidth}
         \begin{center}
          \includegraphics[width=1\linewidth]{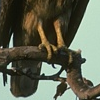}
~
        \end{center}
       \end{minipage}
       \begin{minipage}{0.116\linewidth}
        \begin{center}
          \includegraphics[width=1\linewidth]{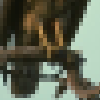}
~
        \end{center}
       \end{minipage}
       \begin{minipage}{0.116\linewidth}
        \begin{center}
          \includegraphics[width=1\linewidth]{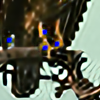}
(17.57)
        \end{center}
      \end{minipage}
       \begin{minipage}{0.116\linewidth}
        \begin{center}
          \includegraphics[width=1\linewidth]{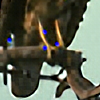}
(20.44)
        \end{center}
       \end{minipage}
       \begin{minipage}{0.116\linewidth}
        \begin{center}
          \includegraphics[width=1\linewidth]{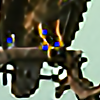}
(17.79)
        \end{center}
       \end{minipage}
       \begin{minipage}{0.116\linewidth}
        \begin{center}
          \includegraphics[width=1\linewidth]{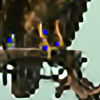}
(19.08)
        \end{center} 
       \end{minipage}
       \begin{minipage}{0.116\linewidth}
        \begin{center}
          \includegraphics[width=1\linewidth]{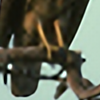}
\blue{(24.76)}
        \end{center} 
       \end{minipage}
       \begin{minipage}{0.116\linewidth}
        \begin{center}
          \includegraphics[width=1\linewidth]{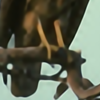}
\red{(25.30)}
        \end{center} 
       \end{minipage}
    \end{minipage}\\
    \begin{minipage}{1\linewidth}
       \begin{minipage}{0.116\linewidth}
         \begin{center}
          \includegraphics[width=1\linewidth]{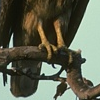}
~
        \end{center}
       \end{minipage}
       \begin{minipage}{0.116\linewidth}
        \begin{center}
          \includegraphics[width=1\linewidth]{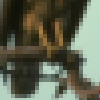}
~
        \end{center}
       \end{minipage}
       \begin{minipage}{0.116\linewidth}
        \begin{center}
          \includegraphics[width=1\linewidth]{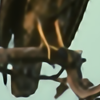}
(24.63)
        \end{center}
      \end{minipage}
       \begin{minipage}{0.116\linewidth}
        \begin{center}
          \includegraphics[width=1\linewidth]{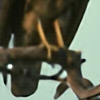}
(25.06)
        \end{center}
       \end{minipage}
       \begin{minipage}{0.116\linewidth}
        \begin{center}
          \includegraphics[width=1\linewidth]{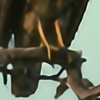}
(24.39)
        \end{center}
       \end{minipage}
       \begin{minipage}{0.116\linewidth}
        \begin{center}
          \includegraphics[width=1\linewidth]{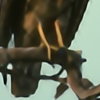}
(24.68)
        \end{center} 
       \end{minipage}
       \begin{minipage}{0.116\linewidth}
        \begin{center}
          \includegraphics[width=1\linewidth]{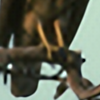}
\blue{(25.36)}
        \end{center} 
       \end{minipage}
       \begin{minipage}{0.116\linewidth}
        \begin{center}
          \includegraphics[width=1\linewidth]{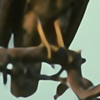}
\red{(26.79)}
        \end{center} 
       \end{minipage}
    \end{minipage}\\
     \begin{minipage}{1\linewidth}
       \begin{minipage}{0.116\linewidth}
         \begin{center}
          \includegraphics[width=1\linewidth]{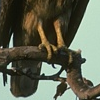}
~
        \end{center}
       \end{minipage}
       \begin{minipage}{0.116\linewidth}
        \begin{center}
          \includegraphics[width=1\linewidth]{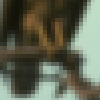}
~
        \end{center}
       \end{minipage}
       \begin{minipage}{0.116\linewidth}
        \begin{center}
          \includegraphics[width=1\linewidth]{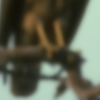}
(20.82)
        \end{center}
      \end{minipage}
       \begin{minipage}{0.116\linewidth}
        \begin{center}
          \includegraphics[width=1\linewidth]{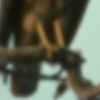}
(21.53)
        \end{center}
       \end{minipage}
       \begin{minipage}{0.116\linewidth}
        \begin{center}
          \includegraphics[width=1\linewidth]{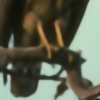}
\blue{(25.03)}
        \end{center}
       \end{minipage}
       \begin{minipage}{0.116\linewidth}
        \begin{center}
          \includegraphics[width=1\linewidth]{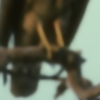}
(22.41)
        \end{center} 
       \end{minipage}
       \begin{minipage}{0.116\linewidth}
        \begin{center}
          \includegraphics[width=1\linewidth]{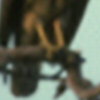}
(21.63)
        \end{center} 
       \end{minipage}
       \begin{minipage}{0.116\linewidth}
        \begin{center}
          \includegraphics[width=1\linewidth]{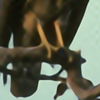}
\red{(26.81)}
        \end{center} 
       \end{minipage}
     \end{minipage}\\
     \begin{minipage}{1\linewidth}
       \begin{minipage}{0.116\linewidth}
         \begin{center}
          \includegraphics[width=1\linewidth]{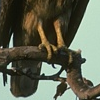}
~\\ \subcaption{HR}
          \label{fig:sigma_40_bsd100_017_hr}
        \end{center}
       \end{minipage}
       \begin{minipage}{0.116\linewidth}
        \begin{center}
          \includegraphics[width=1\linewidth]{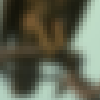}
~\\ \subcaption{LR}
          \label{fig:sigma_40_bsd100_017_input}
        \end{center}
       \end{minipage}
       \begin{minipage}{0.116\linewidth}
        \begin{center}
          \includegraphics[width=1\linewidth]{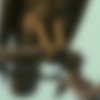}
(18.72)\\ \subcaption{DBPN}
          \label{fig:sigma_40_bsd100_017_dbpn}
        \end{center}
      \end{minipage}
       \begin{minipage}{0.116\linewidth}
        \begin{center}
          \includegraphics[width=1\linewidth]{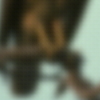}
(18.92)\\ \subcaption{DBPN-Bl}
          \label{fig:sigma_40_bsd100_017_dbpn_bl}
        \end{center}
       \end{minipage}
       \begin{minipage}{0.116\linewidth}
        \begin{center}
          \includegraphics[width=1\linewidth]{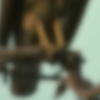}
\blue{(20.15)}\\ \subcaption{IKC}
          \label{fig:sigma_40_bsd100_017_ikc}
        \end{center}
       \end{minipage}
       \begin{minipage}{0.116\linewidth}
        \begin{center}
          \includegraphics[width=1\linewidth]{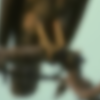}
(19.64)\\ \subcaption{DAN}
          \label{fig:sigma_40_bsd100_017_dan}
        \end{center} 
       \end{minipage}
       \begin{minipage}{0.116\linewidth}
        \begin{center}
          \includegraphics[width=1\linewidth]{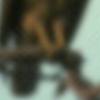}
(19.12)\\ \subcaption{KCBPN}
          \label{fig:sigma_40_bsd100_017_kcbpn}
        \end{center} 
       \end{minipage}
       \begin{minipage}{0.116\linewidth}
        \begin{center}
          \includegraphics[width=1\linewidth]{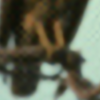}
\red{(21.05)}\\ \subcaption{KBPN}
          \label{fig:sigma_40_bsd100_017_kbpn}
        \end{center} 
      \end{minipage}
    \end{minipage}\\
    \end{tabular}
    \caption{Visual SR comparison. HR images are degraded with the
      isotropic Gaussian kernels ($\sigma \in \{0.2, 1.3, 2.6, 4.0 \}$).  The
      first, second, third and fourth rows show the results of
      $\sigma=0.2$, $1.3$, $2.6$, and $4.0$, respectively.  A value in
      each parenthesis is PSNR.
      The best and the second-best scores are colored by
      \red{red} and \blue{blue}, respectively, in each row.}
    \label{fig:samples}
  \end{center}
\end{figure*}

\begin{figure*}
  \begin{center}
    \begin{tabular}{c}
     \begin{minipage}{1\linewidth}
       \begin{minipage}{0.120\linewidth}
         \begin{center}
          \includegraphics[width=1\linewidth]{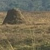}
~
        \end{center}
       \end{minipage}
     \begin{minipage}{0.120\linewidth}
         \begin{center}
          \includegraphics[width=1\linewidth]{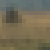}
~
        \end{center}
       \end{minipage}
       \begin{minipage}{0.120\linewidth}
        \begin{center}
          \includegraphics[width=1\linewidth]{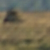}
(28.62, \blue{4.346})
        \end{center}
       \end{minipage}
       \begin{minipage}{0.120\linewidth}
        \begin{center}
          \includegraphics[width=1\linewidth]{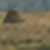}
(\blue{28.92}, 5.212)
        \end{center}
      \end{minipage}
       \begin{minipage}{0.120\linewidth}
        \begin{center}
          \includegraphics[width=1\linewidth]{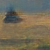}
(27.23, \red{4.022})
        \end{center}
       \end{minipage}
       \begin{minipage}{0.120\linewidth}
        \begin{center}
          \includegraphics[width=1\linewidth]{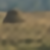}
(28.87, 4.984) %
        \end{center} 
       \end{minipage}
       \begin{minipage}{0.120\linewidth}
        \begin{center}
          \includegraphics[width=1\linewidth]{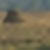}
(28.74, 4.972) %
        \end{center} 
       \end{minipage}
       \begin{minipage}{0.120\linewidth}
        \begin{center}
          \includegraphics[width=1\linewidth]{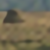}
(\red{28.97}, 5.773)
        \end{center} 
      \end{minipage}
    \end{minipage}\\
     \begin{minipage}{1\linewidth}
       \begin{minipage}{0.120\linewidth}
         \begin{center}
          \includegraphics[width=1\linewidth]{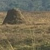}
~\\ 
        \end{center}
       \end{minipage}
     \begin{minipage}{0.120\linewidth}
         \begin{center}
          \includegraphics[width=1\linewidth]{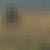}
~
        \end{center}
       \end{minipage}
       \begin{minipage}{0.120\linewidth}
        \begin{center}
          \includegraphics[width=1\linewidth]{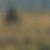}
(27.31, \red{4.175})\\ 
        \end{center}
       \end{minipage}
       \begin{minipage}{0.120\linewidth}
        \begin{center}
          \includegraphics[width=1\linewidth]{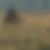}
(27.42, 6.163)\\ 
        \end{center}
      \end{minipage}
       \begin{minipage}{0.120\linewidth}
        \begin{center}
          \includegraphics[width=1\linewidth]{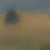}
(26.78, \blue{5.413})\\ 
        \end{center}
       \end{minipage}
       \begin{minipage}{0.120\linewidth}
        \begin{center}
          \includegraphics[width=1\linewidth]{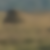}
(\blue{27.73}, 5.833)\\ 
        \end{center} 
       \end{minipage}
       \begin{minipage}{0.120\linewidth}
        \begin{center}
          \includegraphics[width=1\linewidth]{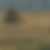}
(27.17, 5.923)\\ 
        \end{center} 
       \end{minipage}
       \begin{minipage}{0.120\linewidth}
        \begin{center}
          \includegraphics[width=1\linewidth]{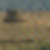}
(\red{28.07}, 5.746)\\ 
        \end{center} 
      \end{minipage}
    \end{minipage}\\
     \begin{minipage}{1\linewidth}
       \begin{minipage}{0.120\linewidth}
         \begin{center}
          \includegraphics[width=1\linewidth]{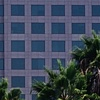}
~
        \end{center}
       \end{minipage}
     \begin{minipage}{0.120\linewidth}
         \begin{center}
          \includegraphics[width=1\linewidth]{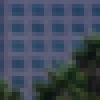}
~
        \end{center}
       \end{minipage}
       \begin{minipage}{0.120\linewidth}
        \begin{center}
          \includegraphics[width=1\linewidth]{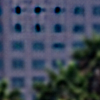}
(23.18, \blue{4.678}) %
        \end{center}
       \end{minipage}
       \begin{minipage}{0.120\linewidth}
        \begin{center}
          \includegraphics[width=1\linewidth]{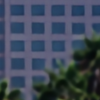}
(24.91, 5.503) %
        \end{center}
      \end{minipage}
       \begin{minipage}{0.120\linewidth}
        \begin{center}
          \includegraphics[width=1\linewidth]{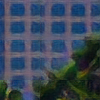}
(23.99, \red{3.721}) %
        \end{center}
       \end{minipage}
       \begin{minipage}{0.120\linewidth}
        \begin{center}
          \includegraphics[width=1\linewidth]{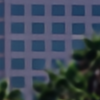}
(\blue{25.71}, 5.445) 
        \end{center} 
       \end{minipage}
       \begin{minipage}{0.120\linewidth}
        \begin{center}
          \includegraphics[width=1\linewidth]{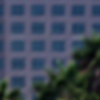}
(24.60, 6.244) 
        \end{center} 
       \end{minipage}
       \begin{minipage}{0.120\linewidth}
        \begin{center}
          \includegraphics[width=1\linewidth]{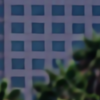}
(\red{25.74}, 6.545) %
        \end{center} 
      \end{minipage}
    \end{minipage}\\
     \begin{minipage}{1\linewidth}
       \begin{minipage}{0.120\linewidth}
         \begin{center}
          \includegraphics[width=1\linewidth]{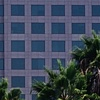}
~\\ 
        \end{center}
       \end{minipage}
     \begin{minipage}{0.120\linewidth}
         \begin{center}
          \includegraphics[width=1\linewidth]{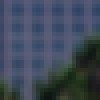}
~
        \end{center}
       \end{minipage}
       \begin{minipage}{0.120\linewidth}
        \begin{center}
          \includegraphics[width=1\linewidth]{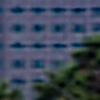}
(23.85, \blue{4.604})\\ 
        \end{center}
       \end{minipage}
       \begin{minipage}{0.120\linewidth}
        \begin{center}
          \includegraphics[width=1\linewidth]{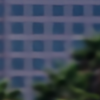}
(\blue{26.42}, 7.321)\\ 
        \end{center} 
      \end{minipage}
       \begin{minipage}{0.120\linewidth}
        \begin{center}
          \includegraphics[width=1\linewidth]{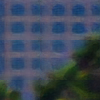}
(25.19, \red{4.106})\\ 
        \end{center}
       \end{minipage}
       \begin{minipage}{0.120\linewidth}
        \begin{center}
          \includegraphics[width=1\linewidth]{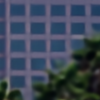}
(26.30, 5.403)\\ 
        \end{center} 
       \end{minipage}
       \begin{minipage}{0.120\linewidth}
        \begin{center}
          \includegraphics[width=1\linewidth]{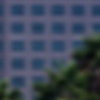}
(25.17, 5.582)\\ 
        \end{center} 
       \end{minipage}
       \begin{minipage}{0.120\linewidth}
        \begin{center}
          \includegraphics[width=1\linewidth]{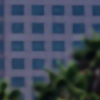}
(\red{26.65}, 7.430)\\ 
        \end{center} 
      \end{minipage}
    \end{minipage}\\
    \begin{minipage}{1\linewidth}
       \begin{minipage}{0.120\linewidth}
         \begin{center}
          \includegraphics[width=1\linewidth]{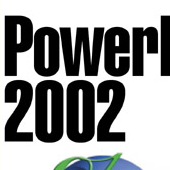}
~\\ 
        \end{center}
       \end{minipage}
     \begin{minipage}{0.120\linewidth}
         \begin{center}
          \includegraphics[width=1\linewidth]{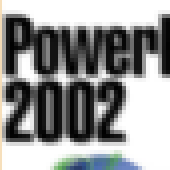}
~
        \end{center}
       \end{minipage}
       \begin{minipage}{0.120\linewidth}
        \begin{center}
          \includegraphics[width=1\linewidth]{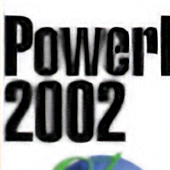}
(19.30, 5.005)\\ 
        \end{center}
       \end{minipage}
       \begin{minipage}{0.120\linewidth}
        \begin{center}
          \includegraphics[width=1\linewidth]{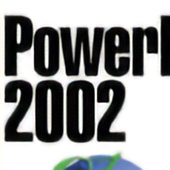}
(25.82, 4.633)\\ 
        \end{center}
      \end{minipage}
       \begin{minipage}{0.120\linewidth}
        \begin{center}
          \includegraphics[width=1\linewidth]{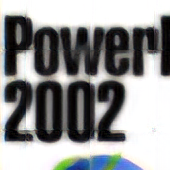}
(16.67, \red{3.083})\\ 
        \end{center}
       \end{minipage}
       \begin{minipage}{0.120\linewidth}
        \begin{center}
          \includegraphics[width=1\linewidth]{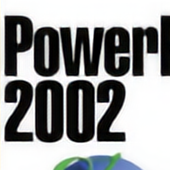}
(25.85, 4.716)\\ 
        \end{center} 
       \end{minipage}
       \begin{minipage}{0.120\linewidth}
        \begin{center}
          \includegraphics[width=1\linewidth]{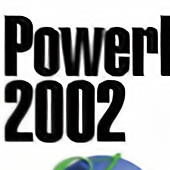}
(\blue{26.13}, 4.728)\\ 
        \end{center} 
       \end{minipage}
       \begin{minipage}{0.120\linewidth}
        \begin{center}
          \includegraphics[width=1\linewidth]{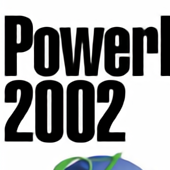}
(\red{33.84},  \blue{4.618})\\ 
        \end{center} 
      \end{minipage}
    \end{minipage}\\
    \begin{minipage}{1\linewidth}
       \begin{minipage}{0.120\linewidth}
         \begin{center}
          \includegraphics[width=1\linewidth]{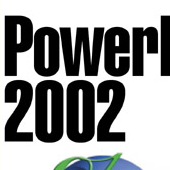}
~
        \end{center}
       \end{minipage}
     \begin{minipage}{0.120\linewidth}
         \begin{center}
          \includegraphics[width=1\linewidth]{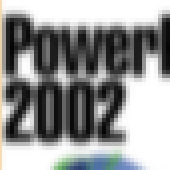}
~
        \end{center}
       \end{minipage}
       \begin{minipage}{0.120\linewidth}
        \begin{center}
          \includegraphics[width=1\linewidth]{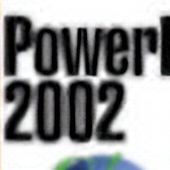}
(15.98, 4.627)
        \end{center}
       \end{minipage}
       \begin{minipage}{0.120\linewidth}
        \begin{center}
          \includegraphics[width=1\linewidth]{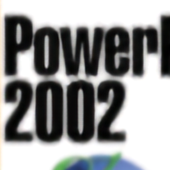}
(18.49, 5.047)
        \end{center}
      \end{minipage}
       \begin{minipage}{0.120\linewidth}
        \begin{center}
          \includegraphics[width=1\linewidth]{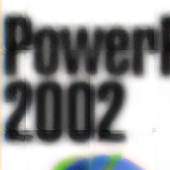}
(13.80, \blue{3.961})
        \end{center}
       \end{minipage}
       \begin{minipage}{0.120\linewidth}
        \begin{center}
          \includegraphics[width=1\linewidth]{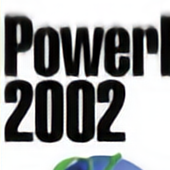}
(\blue{23.49}, 4.415)
        \end{center} 
       \end{minipage}
       \begin{minipage}{0.120\linewidth}
        \begin{center}
          \includegraphics[width=1\linewidth]{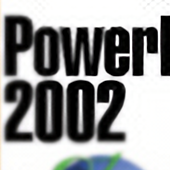}
(20.76, 4.798)
        \end{center} 
       \end{minipage}
       \begin{minipage}{0.120\linewidth}
        \begin{center}
          \includegraphics[width=1\linewidth]{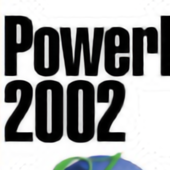}
(\red{26.90}, \red{3.924})
        \end{center} 
      \end{minipage}
    \end{minipage}\\
     \begin{minipage}{1\linewidth}
       \begin{minipage}{0.120\linewidth}
         \begin{center}
          \includegraphics[width=1\linewidth]{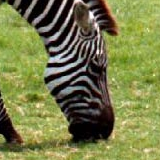}
~
        \end{center}
       \end{minipage}
     \begin{minipage}{0.120\linewidth}
         \begin{center}
          \includegraphics[width=1\linewidth]{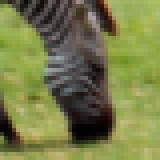}
~
        \end{center}
       \end{minipage}
       \begin{minipage}{0.120\linewidth}
        \begin{center}
          \includegraphics[width=1\linewidth]{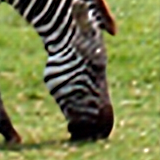}
(19.52, 6.935) 
        \end{center}
       \end{minipage}
       \begin{minipage}{0.120\linewidth}
        \begin{center}
          \includegraphics[width=1\linewidth]{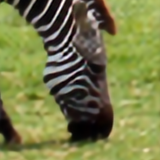}
(19.70, 6.724) 
        \end{center}
      \end{minipage}
       \begin{minipage}{0.120\linewidth}
        \begin{center}
          \includegraphics[width=1\linewidth]{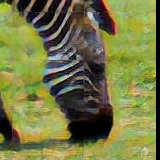}
(17.31, \red{3.906}) 
        \end{center}
       \end{minipage}
       \begin{minipage}{0.120\linewidth}
        \begin{center}
          \includegraphics[width=1\linewidth]{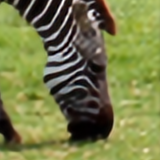}
(\blue{19.74}, 7.009) 
        \end{center} 
       \end{minipage}
       \begin{minipage}{0.120\linewidth}
        \begin{center}
          \includegraphics[width=1\linewidth]{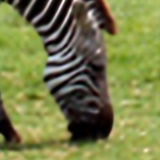}
(\red{20.29}, \blue{6.659}) 
        \end{center} 
       \end{minipage}
       \begin{minipage}{0.120\linewidth}
        \begin{center}
          \includegraphics[width=1\linewidth]{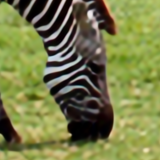}
(19.61, 6.747) 
        \end{center} 
      \end{minipage}
    \end{minipage}\\
     \begin{minipage}{1\linewidth}
       \begin{minipage}{0.120\linewidth}
         \begin{center}
          \includegraphics[width=1\linewidth]{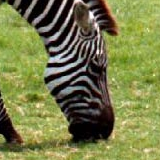}
~\\ \subcaption{HR}
        \end{center}
       \end{minipage}
       \begin{minipage}{0.120\linewidth}
        \begin{center}
          \includegraphics[width=1\linewidth]{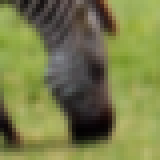}
~\\ \subcaption{LR} %
        \end{center}
       \end{minipage}
       \begin{minipage}{0.120\linewidth}
        \begin{center}
          \includegraphics[width=1\linewidth]{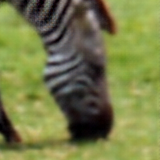}
(19.96, \blue{5.503})\\ \subcaption{DBPN-Bl} %
        \end{center}
       \end{minipage}
       \begin{minipage}{0.120\linewidth}
        \begin{center}
          \includegraphics[width=1\linewidth]{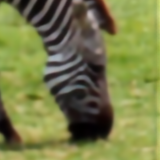}
(\blue{20.25}, 6.159)\\ \subcaption{IKC} 
        \end{center}
      \end{minipage}
       \begin{minipage}{0.120\linewidth}
        \begin{center}
          \includegraphics[width=1\linewidth]{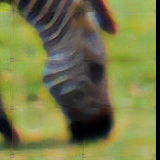}
(17.25, \red{3.870})\\ \subcaption{DASR--Gu} 
        \end{center}
       \end{minipage}
       \begin{minipage}{0.120\linewidth}
        \begin{center}
          \includegraphics[width=1\linewidth]{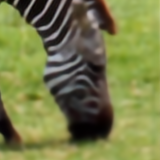}
(20.16, 5.997)\\ \subcaption{DASR--Wang} 
        \end{center} 
       \end{minipage}
       \begin{minipage}{0.120\linewidth}
        \begin{center}
          \includegraphics[width=1\linewidth]{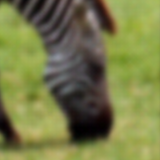}
(19.24, 5.848)\\ \subcaption{KCBPN} 
        \end{center} 
       \end{minipage}
       \begin{minipage}{0.120\linewidth}
        \begin{center}
          \includegraphics[width=1\linewidth]{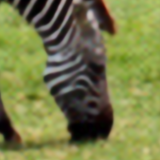}
(\red{20.37}, 6.566)\\ \subcaption{KBPN} 
        \end{center} 
      \end{minipage}
    \end{minipage}\\
    \end{tabular}
    \caption{Visual SR comparison.
      In each example, The first and second rows show the results of test images degraded with $\sigma_{x}/\sigma_{y}=1.3/2.6$
      and $2.6/4.0$, respectively.  Values in each parenthesis are
      (PSNR, NIQE$\downarrow$).
      The best and the second-best scores are colored by \red{red} and \blue{blue},
      respectively, in each row.}
    \label{fig:anisotropic_samples}
  \end{center}
\end{figure*}

\noindent
{\bf Evaluation:}
Set5~\cite{set5}, Set14~\cite{set14}, and BSD100~\cite{BSDS100} are
used for evaluation.
%
%
For experiments shown in Tables~\ref{table:ex_comparison},
\ref{table:ex_comparison_fully_blind},
and \ref{table:ex_ablation_LR_loss}
and Fig.~\ref{fig:samples},
the test images are blurred by the isotropic Gaussian kernels with
$\sigma \in \{0.2, 1.3, \frac{3.5}{\sqrt{2}}, 2.6,
\frac{3.5}{\sqrt{2}}, 4.0 \}$.
The test images used in Table~\ref{table:ex_comparison_anisotropic} and Fig.~\ref{fig:anisotropic_samples} are blurred by anisotropic Gaussians.
%
%
The quantitative evaluation code we used is available in~\cite{pirm2018-github}.
%
%

\noindent
{\bf Codes and trained weights:}
All SOTA methods (i.e., DAN~\cite{DAN-github},
ZSSR~\cite{zssr-github},
Correction Filter~\cite{correction-github},
KOALAnet~\cite{koalanet-github},
DASR--Gu~\cite{dasr-gu-github}, 
DASR--Wang~\cite{dasr-wang-github}, and FAIG~\cite{bib:xie2021nips})
are evaluated with the authors' codes and weights, except that IKC is
evaluated with a publicly-available code~\cite{IKC-github} because its
official code is unavailable.
Since the trained weights given by the authors are used for a fair comparison, comparative methods differ between the results with  isotropic and anisotropic blurs shown in Table~\ref{table:ex_comparison} and Table~\ref{table:ex_comparison_anisotropic}, respectively, depending on which blur is trained for which method.
The scores of (a) and (b) in Table~\ref{table:ex_comparison_fully_blind} are provided in~\cite{hussein2020correction}.
%
%
Since all of these methods including KCBPN and KBPN are designed with reconstruction losses with no perceptual losses, they can be fairly compared.

\begin{table}[t]
  \caption{Ablation studies on KBPN with 7 stages. Both training and test images are blurred by isotropic blur kernels.}
  \label{table:ex_ablation_LR_loss}
  \begin{center}
    \begin{tabular}{l||ccc}\hline
      Method &Set5 &Set14 &BSD100 \\ \hline\hline
      %
      KBPN w/o enhancement & 29.72& 26.79& 26.37\\
      KBPN w/o $L_{LR}$ & 30.20& 27.17& 26.54\\
      KBPN w/o $L_{K}$ & 30.15& 27.01& 26.48\\
      KBPN w/o deconvs & 30.06 & 27.07 & 26.59\\
      KBPN  & \red{30.38} & \red{27.39} & \red{26.70} \\
      \hline    
    \end{tabular}			
  \end{center}
\end{table}

\begin{table*}[t]
  \caption{Comparison with blind SR methods. Both training and test images are blurred by anisotropic blur kernels.
  In addition to PSNR and SSIM, NIQE and NRQM are also
    evaluated as perceptual quality metrics.}
  \label{table:ex_comparison_anisotropic}
  \begin{center}
  \begin{footnotesize}
    \begin{tabular}{l||c|cccc|cccc|cccc}\hline
      \multirow{2}{*}{Method} &
      \multirow{2}{*}{\hspace*{-1mm}$\sigma_{x}/\sigma_{y}$\hspace*{-1mm}}& \multicolumn{4}{|c}{Set5}&\multicolumn{4}{|c}{Set14}&\multicolumn{4}{|c}{BSD100}\\
      &&PSNR&SSIM  &NIQE$\downarrow$&NRQM  &PSNR&SSIM & NIQE$\downarrow$&NRQM  &PSNR&SSIM  &NIQE$\downarrow$&NRQM\\ \hline\hline

    DBPN~\cite{DBLP:journals/pami/HarisSU21} &\multirow{10}{*}{\hspace*{-1mm}1.3/2.6\hspace*{-1mm}} & 29.02 & 0.834 & 10.932 & 3.963 & 26.36 & 0.721 & 7.283 & 4.232 & 26.34 & 0.686 & 7.466 & 4.032\\
      DBPN-Bl                               & & 29.83 & 0.848 & \blue{7.249} & 4.637 & 26.45 & 0.725 & \blue{6.070} & 4.397 & 26.57 & 0.696 & \blue{6.298} & 4.560\\
      IKC~\cite{DBLP:conf/cvpr/GuLZD19}     & & 30.14 & 0.861 & 8.227 & 4.932 & 27.08 & 0.742 & 6.752 & 4.761 & 26.93 & 0.713 & 7.161 & \green{5.067}\\
      KOALAnet~\cite{DBLP:conf/cvpr/KimSK21}& & 30.49 & 0.866 & 9.227 & 4.387 & 26.98 & 0.735 & 7.209 & 4.216 & 26.80 & 0.702 & 7.357 & 4.188\\
      DASR--Gu~\cite{DBLP:conf/cvpr/WeiGLTJS21}    & & 22.90 & 0.729 & \red{6.323} & \red{6.923} & 22.85 & 0.632 & \red{4.402} & \red{7.137} & 25.23 & 0.620 & \red{4.565} & \red{7.456}\\
      DASR--Wang~\cite{DBLP:conf/cvpr/WangWDX0AG21}& & 30.66 & 0.869 & 8.406 & 4.831 & 27.00 & 0.737 & 7.035 & 4.491 & 26.90 & 0.707 & 7.221 & 4.607\\
      FAIG~\cite{bib:xie2021nips} & & 30.02 & 0.866 & 10.263 & \blue{5.227} & 27.11 & 0.744 & 7.071 & 4.641 & 26.81 & 0.717 & 7.199 & 5.033 \\
      KCBPN (stg4) & & 30.02 & 0.854 & 8.481 & 4.340 & 27.35 & 0.747 & 7.223 & 4.385 & 26.48 & 0.689 & 7.563 & 4.039\\
      KBPN (stg3)  & & \green{31.55} & \green{0.882} & 8.712 & 4.949 & \blue{28.26} & \green{0.771} & 6.786 & 4.756 & \green{27.32} & \green{0.726} & 7.209 & 4.961\\
      KBPN (stg4)  & & \red{31.99} & \red{0.888} & 8.265 & 4.983 & \red{28.55} & \red{0.780} & \green{6.344} & \blue{5.209} & \red{27.47} & \red{0.731} & \green{6.971} & \blue{5.203}\\
      KBPN (stg7)  & & \blue{31.58} & \blue{0.884} & \green{8.086} & \green{5.004} & \green{28.25} & \blue{0.773} & 6.657 & \green{4.856} & \blue{27.35} & \blue{0.728} & 7.176 & 4.995\\
      \hline
      DBPN~\cite{DBLP:journals/pami/HarisSU21} &\multirow{10}{*}{\hspace*{-1mm}2.6/4.0\hspace*{-1mm}} & 25.40 & 0.721 & 11.192 & 2.499 & 23.79 & 0.607 & 10.036 & 2.346 & 24.29 & 0.587 & 9.922 & 2.397\\
      DBPN-Bl                               & & 27.11 & 0.772 & 8.376 & 3.443 & 24.93 & 0.650 & \blue{7.384} & 3.502 & 25.08 & 0.619 & \blue{7.567} & 3.340\\
      IKC~\cite{DBLP:conf/cvpr/GuLZD19}     & & 27.95 & 0.809 & \green{8.361} & \red{4.415} & 25.54 & 0.676 & \green{7.511} & 3.740 & 25.79 & \green{0.659} & 7.865 & \green{3.810}\\
      KOALAnet~\cite{DBLP:conf/cvpr/KimSK21}& & 27.40 & 0.790 & 10.384 & 3.374 & 25.09 & 0.658 & 8.347 & 3.301 & 25.27 & 0.630 & 8.533 & 3.150\\
      DASR--Gu~\cite{DBLP:conf/cvpr/WeiGLTJS21}    & & 21.91 & 0.682 & \red{6.883} & 3.688 & 21.96 & 0.583 & \red{5.916} & 3.397 & 24.37 & 0.578 & \red{5.950} & 3.601\\
      DASR--Wang~\cite{DBLP:conf/cvpr/WangWDX0AG21}& & \blue{28.93} & \red{0.827} & \blue{8.342} & 3.911 & 25.96 & \green{0.688} & 7.512 & 3.659 & \red{26.16} & \red{0.668} & 7.645 & 3.744\\
      FAIG~\cite{bib:xie2021nips} & & 26.57 & 0.761 & 11.444 & 2.945 & 24.54 & 0.635 & 9.213 & 2.763 & 24.88 & 0.615 & 9.256 & 2.821 \\
      KCBPN (stg4) & & 25.85 & 0.732 & 10.174 & 2.768 & 24.48 & 0.636 & 8.853 & 2.849 & 24.58 & 0.598 & 9.171 & 2.620\\
      KBPN (stg3) & & \red{29.20} & \blue{0.825} & 10.425 & \blue{4.172} & \blue{26.54} & \red{0.704} & 7.563 & \blue{3.971} & \blue{26.09} & \blue{0.662} & 7.636 & \blue{3.950}\\
      KBPN (stg4) & & 28.30 & 0.809 & 13.869 & 3.820 & \green{26.44} & \blue{0.703} & 7.783 & \green{3.831} & 25.87 & 0.653 & 7.851 & 3.700\\
      KBPN (stg7)  & & \green{28.82} & \green{0.815} & 10.103 & \green{4.076} & \red{26.58} & \red{0.704} & 7.526 & \red{4.031} & \green{26.07} & \blue{0.662} & \green{7.592} & \red{4.024}\\
      \hline
    \end{tabular}
  \end{footnotesize}
\end{center}
\end{table*}

\subsection{Quantitative Evaluation}
\label{subsection:results}

\noindent
{\bf Isotropic blur kernels:}
Table~\ref{table:ex_comparison} shows PSNR and SSIM of blind SR methods.
As the baselines, a non-kernel SR method (DBPN~\cite{DBLP:journals/pami/HarisSU21}) and its extension trained with various blur images (i.e., DBPN-Bl) are shown.
Since KCBPN and KBPN outperform DBPN-Bl, which has no kernel representation, in all of $\sigma= \{ 0.2, 1.3, 2.6, 4.0 \}$, the effectiveness of the kernel representation is validated.
DBPN-Bl, KCBPN, and KBPN are superior to IKC and DAN in $\sigma=0.2$.
In larger blurs (i.e., $\sigma= \{ 2.6, 4.0 \}$), IKC and DAN are better than DBPN-Bl and KCBPN in most cases.
However, KBPN is the best in all datasets, metrics, and blurs except that KBPN is the second best in four cases (i.e., PSNR and SSIM of BSD100 in $\sigma=2.6$, SSIM of Set5 in $\sigma=4.0$, and SSIM of BSD100 in $\sigma=4.0$, while the gaps from the best ones are small in all of these four cases.

\noindent
{\bf vs blind image-specific SR:}
Table~\ref{table:ex_comparison_fully_blind}\footnote{$\sigma = \{ \frac{3.5}{\sqrt{2}}, \frac{4.5}{\sqrt{2}} \}$ are selected in accordance with~\cite{hussein2020correction}.} shows a comparison with blind image-specific methods, each of which is self-supervised by a test image.
Kernel GAN~\cite{kernelgan} and Correction
filter~\cite{hussein2020correction} are
blind SR that requires an extra SR network. As the extra SR networks,
ZSSR~\cite{ZSSR} and DBPN~\cite{haris2018deep} were used.
%

It can be seen that KBPN outperforms the image-specific blind SR methods in all datasets, metrics, and blurs, except that KBPN is the second best in two cases (i.e., SSIM of BSD100 in $\sigma=\frac{3.5}{\sqrt{2}}$ and $\sigma=\frac{4.5}{\sqrt{2}}$) in
Table~\ref{table:ex_comparison_fully_blind}.
Furthermore, larger PSNR gains from the other methods are observed with a larger $\sigma$ in Table~\ref{table:ex_comparison_fully_blind} as well as in Table~\ref{table:ex_comparison}.

\noindent
{\bf Ablation study in KBPN:}
As demonstrated in Tables~\ref{table:ex_comparison} and \ref{table:ex_comparison_fully_blind}, our main contributions exist in KBPN.
In KBPN, our proposed SR feature enhancement and LR loss, $\mathcal{L}_{LR}$, are distinct components compared with other SR methods.
In addition, we also evaluate the effect of the kernel loss, $\mathcal{L}_{K}$.
Furthermore, the deconv layers in the connection  from the blur branch to the SR branch are ablated.
This ablation is done by bringing the destination of this connection to the path between $\acute{F}^{LR}_{1,...,t}$ and SFT (see Fig.~\ref{fig:kbpn}).
The effects of these three components are verified in the ablation study.

The effects of these four components are shown in Table~\ref{table:ex_ablation_LR_loss}.
The PSNR scores are the mean values of $\sigma = \{0.2,1.3,2.6,4.0\}$ of all the datasets.
We can see that all of the SR feature enhancement, $L_{LR}, L_{K}$, and the deconv layers can improve the SR performance in all cases in Table~\ref{table:ex_ablation_LR_loss}.

\noindent
{\bf Anisotropic blur kernels:}
Blind SR methods trained by anisotropically-blurred images are shown in Table~\ref{table:ex_comparison_anisotropic}.
In addition to PSNR and SSIM, two perceptual scores (i.e., NIQE~\cite{DBLP:journals/spl/MittalSB13} and NRQM~\cite{DBLP:journals/cviu/MaYY017}) are assessed.
Unlike other metrics, a lower value is better in NIQE.
Different from the results of isotropically-blurred images,
it is difficult to choose a method that outperforms all others in all metrics.
We can see the following observations:
\begin{itemize}
\item
In the distortion scores (i.e., PSNR and SSIM), either of KBPNs with the different numbers of stages is the best in most cases; only in PSNR and SSIM of BSD100 in $\sigma_{x}/\sigma_{y} = 2.6/4.0$, DASR--Wang is the best.
The second and third bests are also observed in KBPN.
\item
In the perceptual scores, DASR--Gu~\cite{DBLP:conf/cvpr/WeiGLTJS21} is the best in a smaller blur (i.e., $\sigma_{x}/\sigma_{y} = 1.3/2.6$), while KBPN follows DASR--Gu so that KBPN with four stages gets two second-best scores and two third-best scores.
In a larger blur, KBPN with seven stages gets two best scores and two third-best scores, while DASR--Gu is the best in three cases.
While the perceptual scores of DASR--Gu are better in many cases, its distortion scores (PSNR and SSIM) are much worse than KBPN; for example, the PSNR scores of DASR--Gu and KBPN with four stages are 23.20 and 28.10, respectively, on average over all cases shown in Table~\ref{table:ex_comparison_anisotropic}.
\end{itemize}

Based on the above observations, we can conclude that KBPN is the best in terms of the trade-off between the distortion and perceptual scores, especially in a larger blur, as with in Table~\ref{table:ex_comparison}.

\begin{table}[t]
 \caption{Computational cost: the number of parameters (Mega) 
 }
  \label{table:cost}
  \begin{center}
  \begin{tabular}{l|cccc}
    \hline
    & KCBPN (stg4) & KBPN (stg4) & IKC & KOALAnet \\ \hline
    Params & 8.0 & 61.2 & 9.1  & 6.5 \\ \hline
    & AMNet & DAN & DASR--Gu & DASR-Wang \\ \hline
    Params & 22.0 & 4.3  & 16.7 & 6.0  \\ \hline
  \end{tabular}
  \end{center}
\end{table}

\begin{figure}
  \begin{center}
    \includegraphics[width=1\linewidth]{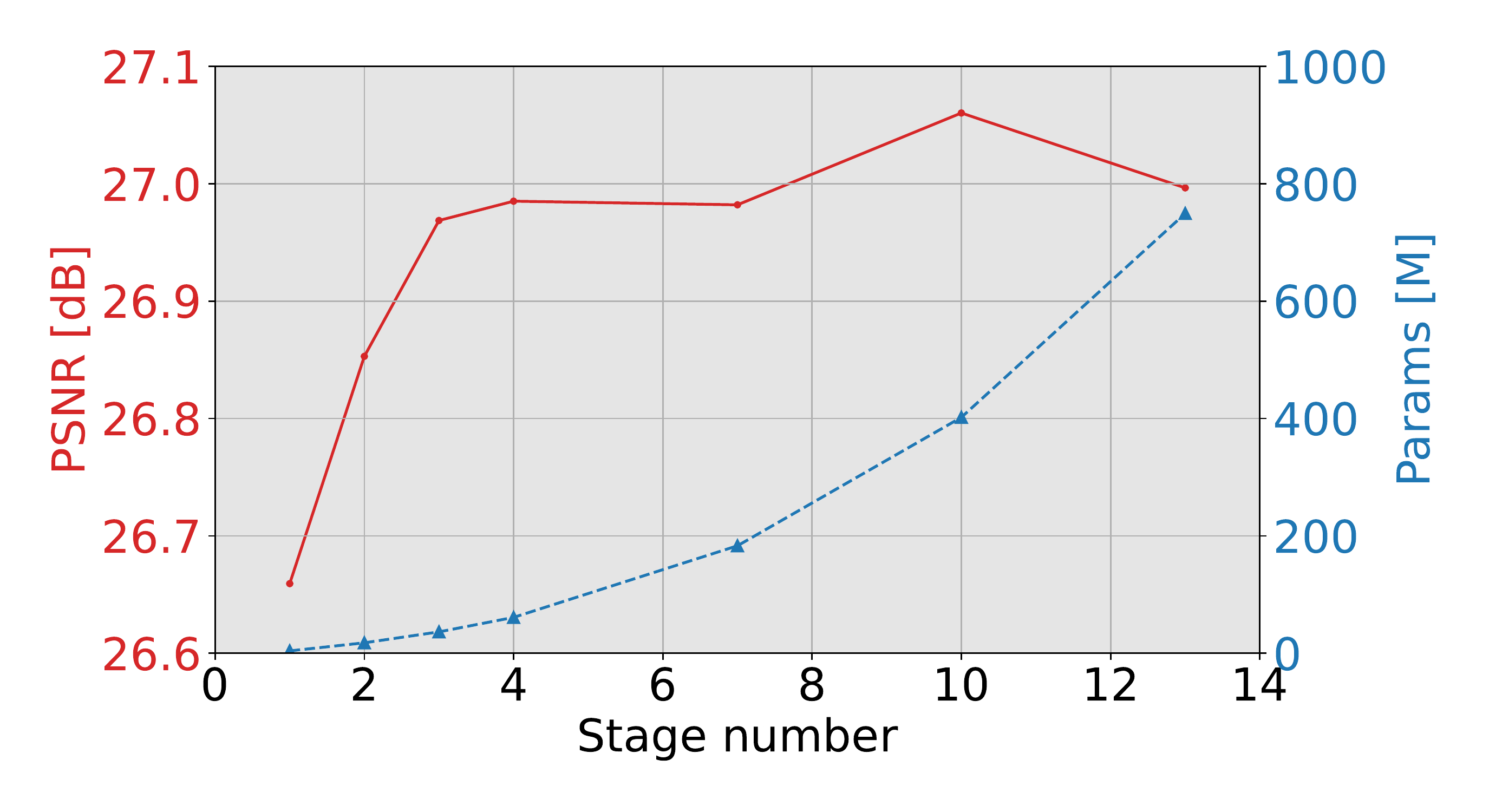}
    \vspace{-1.5em}
    \caption{Relationship between PSNR and the number of parameters changed according to the number of stages in KBPN. The horizontal axis indicates the number of stages. The left and right vertical axes (which correspond to red and blue line graphs, respectively) indicate PSNR and the number of parameters, respectively. These PSNR scores are the mean scores with anisotropic blurs of $\sigma_{x} \in \{0.2,1.3,2.6,4.0\}$ and $\sigma_{y} \in \{0.2,1.3,2.6,4.0\}$ on the BSD100 dataset.}
    \label{fig:number_params}
    \vspace{-1.5em}
  \end{center}
\end{figure}

\noindent
{\bf Computational Cost:}
While iterative stages in KBPN overall improve the SR quality, their computational cost is .
For verifying this limitation compared with other blind SR methods, Table~\ref{table:cost} shows the number of parameters large\footnote{While the iterative process in IKC and DAN is achieved by recurrently employing a single module, KBPN consists of multiple sequentially-connected modules so that the number of the iterations is equal to the one of the modules. Since the single module is shared among all the iteration processes in IKC and DAN, the module size can be reduced compared with the computational cost. KBPN, on the other hand, consits of multiple sequentially-connected modules for improving the performance with the large network capacity.}. 


The relationship between the SR performance and the number of stages (denoted by $N_{s}$) in KBPN is shown in Fig.~\ref{fig:number_params}.
While the number of parameters becomes bigger as $N_{s}$ increases, the SR performance (i.e., PSNR) is almost saturated at $N_{s}=3$.
In particular, the PSNR score decreases after the score is the largest at $N_{s}=10$ probably due to overfitting.


\begin{figure*}[t]
  \begin{center}
    \begin{tabular}{c}
      \begin{minipage}{0.125\linewidth}
         \begin{center}
           \includegraphics[width=1\linewidth]{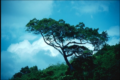}
        \end{center}
       \end{minipage}
      \begin{minipage}{0.050\linewidth}
        \begin{center}
           \includegraphics[width=1\linewidth]{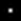}
        \end{center}
      \end{minipage}
      \begin{minipage}{0.050\linewidth}
        \begin{center}
           \includegraphics[width=1\linewidth]{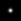}
        \end{center}
      \end{minipage}
      \begin{minipage}{0.125\linewidth}
         \begin{center}
           \includegraphics[width=1\linewidth]{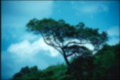}
        \end{center}
       \end{minipage}
      \begin{minipage}{0.050\linewidth}
        \begin{center}
           \includegraphics[width=1\linewidth]{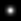}
        \end{center}
      \end{minipage}
      \begin{minipage}{0.050\linewidth}
        \begin{center}
           \includegraphics[width=1\linewidth]{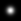}
        \end{center}
      \end{minipage}
      \begin{minipage}{0.125\linewidth}
         \begin{center}
           \includegraphics[width=1\linewidth]{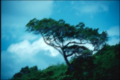}
        \end{center}
       \end{minipage}
      \begin{minipage}{0.050\linewidth}
        \begin{center}
           \includegraphics[width=1\linewidth]{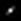}
        \end{center}
      \end{minipage}
      \begin{minipage}{0.050\linewidth}
        \begin{center}
           \includegraphics[width=1\linewidth]{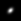}
        \end{center}
      \end{minipage}
      \begin{minipage}{0.125\linewidth}
         \begin{center}
           \includegraphics[width=1\linewidth]{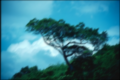}
        \end{center}
       \end{minipage}
      \begin{minipage}{0.050\linewidth}
        \begin{center}
           \includegraphics[width=1\linewidth]{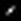}
        \end{center}
      \end{minipage}
      \begin{minipage}{0.050\linewidth}
        \begin{center}
           \includegraphics[width=1\linewidth]{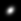}
        \end{center}
      \end{minipage}
      \\
      \vspace{-0.6em}
      \\
 
      \begin{minipage}{0.125\linewidth}
         \begin{center}
           \includegraphics[width=1\linewidth]{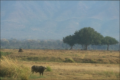}
        \end{center}
       \end{minipage}
      \begin{minipage}{0.050\linewidth}
        \begin{center}
           \includegraphics[width=1\linewidth]{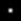}
        \end{center}
      \end{minipage}
      \begin{minipage}{0.050\linewidth}
        \begin{center}
           \includegraphics[width=1\linewidth]{./imgs/Kernel_visualization/KBPN_kondo-stg7-area-iso-2023/sigma_1.3/GT_kernel.png}
        \end{center}
      \end{minipage}
      \begin{minipage}{0.125\linewidth}
         \begin{center}
           \includegraphics[width=1\linewidth]{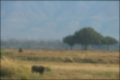}
        \end{center}
       \end{minipage}
      \begin{minipage}{0.050\linewidth}
        \begin{center}
           \includegraphics[width=1\linewidth]{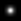}
        \end{center}
      \end{minipage}
      \begin{minipage}{0.050\linewidth}
        \begin{center}
           \includegraphics[width=1\linewidth]{./imgs/Kernel_visualization/KBPN_kondo-stg7-area-iso-2023/sigma_2.6/GT_kernel.png}
        \end{center}
      \end{minipage}
      \begin{minipage}{0.125\linewidth}
         \begin{center}
           \includegraphics[width=1\linewidth]{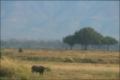}
        \end{center}
       \end{minipage}
      \begin{minipage}{0.050\linewidth}
        \begin{center}
           \includegraphics[width=1\linewidth]{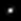}
        \end{center}
      \end{minipage}
      \begin{minipage}{0.050\linewidth}
        \begin{center}
           \includegraphics[width=1\linewidth]{./imgs/Kernel_visualization/KBPN_kondo-stg4-area-seed10-2023/1.3_2.6_45/GT_kernel.png}
        \end{center}
      \end{minipage}
      \begin{minipage}{0.125\linewidth}
         \begin{center}
           \includegraphics[width=1\linewidth]{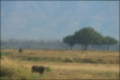}
        \end{center}
       \end{minipage}
      \begin{minipage}{0.050\linewidth}
        \begin{center}
           \includegraphics[width=1\linewidth]{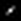}
        \end{center}
      \end{minipage}
      \begin{minipage}{0.050\linewidth}
        \begin{center}
           \includegraphics[width=1\linewidth]{./imgs/Kernel_visualization/KBPN_kondo-stg4-area-seed10-2023/1.3_4.0_45/GT_kernel.png}
        \end{center}
      \end{minipage}
      \\
      \vspace{-0.6em}
      \\
      
      \begin{minipage}{0.125\linewidth}
         \begin{center}
           \includegraphics[width=1\linewidth]{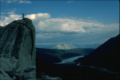}
        \end{center}
       \end{minipage}
      \begin{minipage}{0.050\linewidth}
        \begin{center}
           \includegraphics[width=1\linewidth]{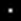}
        \end{center}
      \end{minipage}
      \begin{minipage}{0.050\linewidth}
        \begin{center}
           \includegraphics[width=1\linewidth]{./imgs/Kernel_visualization/KBPN_kondo-stg7-area-iso-2023/sigma_1.3/GT_kernel.png}
        \end{center}
      \end{minipage}
      \begin{minipage}{0.125\linewidth}
         \begin{center}
           \includegraphics[width=1\linewidth]{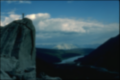}
        \end{center}
       \end{minipage}
      \begin{minipage}{0.050\linewidth}
        \begin{center}
           \includegraphics[width=1\linewidth]{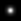}
        \end{center}
      \end{minipage}
      \begin{minipage}{0.050\linewidth}
        \begin{center}
           \includegraphics[width=1\linewidth]{./imgs/Kernel_visualization/KBPN_kondo-stg7-area-iso-2023/sigma_2.6/GT_kernel.png}
        \end{center}
      \end{minipage}
      \begin{minipage}{0.125\linewidth}
         \begin{center}
           \includegraphics[width=1\linewidth]{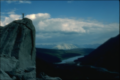}
        \end{center}
       \end{minipage}
      \begin{minipage}{0.050\linewidth}
        \begin{center}
           \includegraphics[width=1\linewidth]{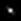}
        \end{center}
      \end{minipage}
      \begin{minipage}{0.050\linewidth}
        \begin{center}
           \includegraphics[width=1\linewidth]{./imgs/Kernel_visualization/KBPN_kondo-stg4-area-seed10-2023/1.3_2.6_45/GT_kernel.png}
        \end{center}
      \end{minipage}
      \begin{minipage}{0.125\linewidth}
         \begin{center}
           \includegraphics[width=1\linewidth]{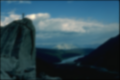}
        \end{center}
       \end{minipage}
      \begin{minipage}{0.050\linewidth}
        \begin{center}
           \includegraphics[width=1\linewidth]{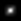}
        \end{center}
      \end{minipage}
      \begin{minipage}{0.050\linewidth}
        \begin{center}
           \includegraphics[width=1\linewidth]{./imgs/Kernel_visualization/KBPN_kondo-stg4-area-seed10-2023/1.3_4.0_45/GT_kernel.png}
        \end{center}
      \end{minipage}
      \\
      \vspace{-0.6em}
      \\
      \begin{minipage}{0.125\linewidth}
         \begin{center}
           \includegraphics[width=1\linewidth]{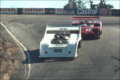}
        \end{center}
       \end{minipage}
      \begin{minipage}{0.050\linewidth}
        \begin{center}
           \includegraphics[width=1\linewidth]{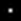}
        \end{center}
      \end{minipage}
      \begin{minipage}{0.050\linewidth}
        \begin{center}
           \includegraphics[width=1\linewidth]{./imgs/Kernel_visualization/KBPN_kondo-stg7-area-iso-2023/sigma_1.3/GT_kernel.png}
        \end{center}
      \end{minipage}
      \begin{minipage}{0.125\linewidth}
         \begin{center}
           \includegraphics[width=1\linewidth]{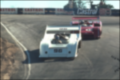}
        \end{center}
       \end{minipage}
      \begin{minipage}{0.050\linewidth}
        \begin{center}
           \includegraphics[width=1\linewidth]{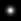}
        \end{center}
      \end{minipage}
      \begin{minipage}{0.050\linewidth}
        \begin{center}
           \includegraphics[width=1\linewidth]{./imgs/Kernel_visualization/KBPN_kondo-stg7-area-iso-2023/sigma_2.6/GT_kernel.png}
        \end{center}
      \end{minipage}
      \begin{minipage}{0.125\linewidth}
         \begin{center}
           \includegraphics[width=1\linewidth]{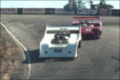}
        \end{center}
       \end{minipage}
      \begin{minipage}{0.050\linewidth}
        \begin{center}
           \includegraphics[width=1\linewidth]{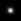}
        \end{center}
      \end{minipage}
      \begin{minipage}{0.050\linewidth}
        \begin{center}
           \includegraphics[width=1\linewidth]{./imgs/Kernel_visualization/KBPN_kondo-stg4-area-seed10-2023/1.3_2.6_45/GT_kernel.png}
        \end{center}
      \end{minipage}
      \begin{minipage}{0.125\linewidth}
         \begin{center}
           \includegraphics[width=1\linewidth]{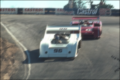}
        \end{center}
       \end{minipage}
      \begin{minipage}{0.050\linewidth}
        \begin{center}
           \includegraphics[width=1\linewidth]{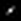}
        \end{center}
      \end{minipage}
      \begin{minipage}{0.050\linewidth}
        \begin{center}
           \includegraphics[width=1\linewidth]{./imgs/Kernel_visualization/KBPN_kondo-stg4-area-seed10-2023/1.3_4.0_45/GT_kernel.png}
        \end{center}
      \end{minipage}
      \\
      \begin{minipage}{0.125\linewidth}
        \begin{center}
        {\scriptsize LR image}
        \end{center}
      \end{minipage}
      \begin{minipage}{0.050\linewidth}
        \begin{center}
        {\scriptsize $K$}
        \end{center}
      \end{minipage}
      \begin{minipage}{0.050\linewidth}
        \begin{center}
        {\scriptsize $K$ (GT)}
        \end{center}
      \end{minipage}
      \begin{minipage}{0.125\linewidth}
         \begin{center}
         {\scriptsize LR image}
        \end{center}
       \end{minipage}
      \begin{minipage}{0.050\linewidth}
        \begin{center}
        {\scriptsize $K$}
        \end{center}
      \end{minipage}
      \begin{minipage}{0.050\linewidth}
        \begin{center}
        {\scriptsize $K$ (GT)}
        \end{center}
      \end{minipage}
      \begin{minipage}{0.125\linewidth}
        \begin{center}
        {\scriptsize LR image}
        \end{center}
       \end{minipage}
      \begin{minipage}{0.050\linewidth}
        \begin{center}
        {\scriptsize $K$}
        \end{center}
      \end{minipage}
      \begin{minipage}{0.050\linewidth}
        \begin{center}
        {\scriptsize $K$ (GT)}
        \end{center}
      \end{minipage}
      \begin{minipage}{0.125\linewidth}
        \begin{center}
        {\scriptsize LR image}
        \end{center}
       \end{minipage}
       \begin{minipage}{0.050\linewidth}
        \begin{center}
        {\scriptsize $K$}
        \end{center}
      \end{minipage}
      \begin{minipage}{0.050\linewidth}
        \begin{center}
        {\scriptsize $K$ (GT)}
        \end{center}
      \end{minipage}
      \\
      \begin{minipage}{0.24\linewidth}
        \begin{center}
        \subcaption{$\sigma = 1.3$}
        \end{center}
       \end{minipage}
      \begin{minipage}{0.24\linewidth}
        \begin{center}
        \subcaption{$\sigma = 2.6$}
        \end{center}
      \end{minipage}
      \begin{minipage}{0.24\linewidth}
         \begin{center}
         \subcaption{$\sigma_{x}/\sigma_{y} = 1.3 / 2.6$}
        \end{center}
       \end{minipage}
      \begin{minipage}{0.24\linewidth}
        \begin{center}
        \subcaption{$\sigma_{x}/\sigma_{y} = 1.3 / 4.0$}
        \end{center}
      \end{minipage}
    \end{tabular}
    \vspace{-1em}
    \caption{Visualization of estimated kernels ($K$) in KBPN. The ground-truth kernels are shown in the rightmost column in each of (a), (b), (c), and (d).
      Note that these LR images are difficult examples for kernel estimation because less textures are captured so that textureless sky and ground are widely observed.
    }
    \label{fig:kernel_samples}
  \end{center}
  \begin{center}
    \includegraphics[width=\linewidth]{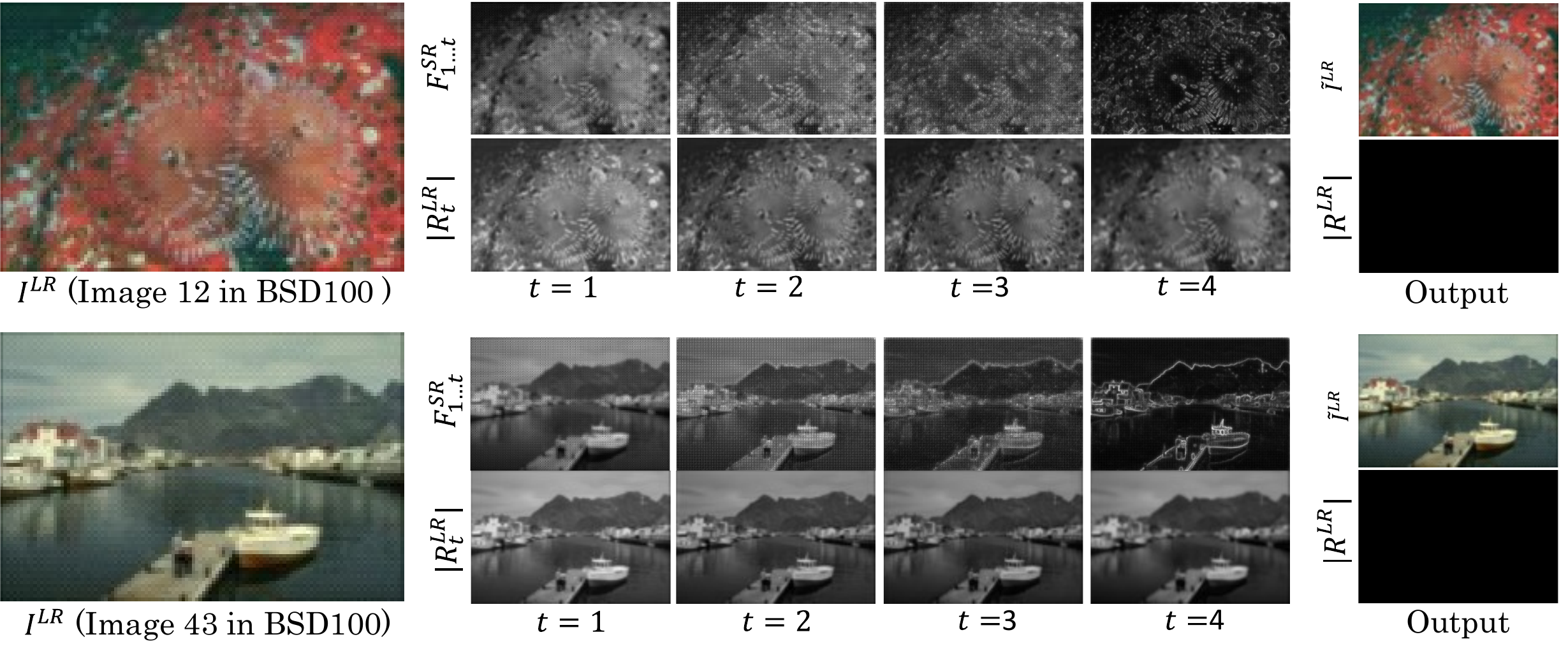}
     \vspace{-1.5em}
    \caption{Visualization results of the SR feature maps $F^{SR}_{1,\cdots,t}$ and the LR image residuals $| R^{LR}_{t} |$ of in KBPN with 4 stages. 
    For reference, the LR image residual computed with the output SR image $I^{SR}$ (i.e., $| R^{LR} | = | \tilde{I}^{LR} - I^{LR} |$ where $\tilde{I}^{LR} = (I^{SR} \ast K)\downarrow_{s}$) is also shown in the lower row of the rightmost column. $I^{LR}$ in this figure are generated with $\sigma_{x}/\sigma_{y}=1.3/2.6$.
    }
    \label{fig:latent_representation}
    \vspace{-1em}
  \end{center}
\end{figure*}

\subsection{Visual Results with Images Blurred by Isotropic Blurs}

Figure~\ref{fig:samples} shows the SR results of KCBPN and KBPN in which both training and test images are isotropically blurred.
All images in Fig.~\ref{fig:samples} are cropped from their original SR images for detailed observations.
For comparison, the results of DBPN~\cite{haris2018deep} as non-blind SR, its extension trained by various blur kernels (DBPN-Bl) as non-kernel SR, IKC~\cite{DBLP:conf/cvpr/GuLZD19}, and DAN~\cite{DBLP:conf/nips/Luo0LWT20}
are shown.
The PSNR score of each SR image is also shown in this figure.
In most examples, KCBPN and KBPN can reconstruct better images as  validated qualitatively and quantitatively.
While other SOTA methods (IKC and DAN) reconstruct sharper boundaries, their PSNR scores are much worse than KCBPN and KBPN because the boundaries reconstructed by IKC and DAN are unrealistically jagged.


\subsection{Visual Results with Images Blurred by Anisotropic Blurs}

Figure~\ref{fig:anisotropic_samples} shows the SR images of KCBPN and KBPN reconstructed
from LR images degraded by anisotropic Gaussian kernels.
The PSNR and NIQE scores are shown with each SR result.
For comparison, DBPN-Bl, IKC~\cite{DBLP:conf/cvpr/GuLZD19}, 
DASR--Gu~\cite{DBLP:conf/cvpr/WeiGLTJS21}, and 
DASR--Wang~\cite{DBLP:conf/cvpr/WangWDX0AG21} are also shown.
%
It can be seen that KBPN is superior to the other methods in terms of the trade-off between PSNR (i.e., image distortion) and NIQE (i.e., perceptual quality), and KBPN tended to emphasize scene edges more clearly, which is an important property in terms of actual perceptual quality.


\subsection{Visualization of Estimated Kernels}

Figure~\ref{fig:kernel_samples} shows examples of blur kernels estimated in test images blurred by isotropic Gaussian kernels ($\sigma = 1.3, 2.6$) and anisotropic Gaussian kernels ($\sigma_{x}/\sigma_{y} = 1.3 / 2.6, 1.3 / 4.0$).
These sample test images contain both complex and less textured images.
Since such complex and less textures make kernel estimation less and more difficult, respectively, these images are appropriate for understanding the pros and cons of our proposed method.
The kernels estimated with more complex anisotropic Gaussian blurs ($\sigma_{x}/\sigma_{y} = 1.3 / 4.0$) are directionally correct, but the estimated variance is not accurate.
More improvement of this kernel estimation is important for future work.
By enhancing the kernel estimation more, it is expected that the quality of KBPN for the anisotropic Gaussian blurs shown in Table~\ref{table:ex_comparison_fully_blind} can be improved more.


\subsection{Feature Representation Analysis}

In~\cite{DBLP:journals/pami/HarisSU21}, feature maps extracted in different iterative stages are visualized for demonstrating that a variety of features can be extracted.
Such feature maps are visualized also in our KBPN for verifying whether or not our iterative stages with the proposed SR feature enhancement can extract various features.
SR feature maps $F^{SR}_{1...t}$ and LR residual maps $R^{LR}_{t}$ in different stages are shown in Fig.~\ref{fig:latent_representation}.
Each feature map is converted to a grayscale image for visualization.

We can see that KBPN extracts a variety of feature maps in $F^{SR}_{1...t}$. 
The first stage tends to represent low-frequency features, while the last stage represents high-frequency features.

In $R^{LR}_{t} = \tilde{I}^{LR}_{t} - I^{LR}$ where $\tilde{I}^{LR}_{t} = (I^{SR}_{t} \ast K_{t})\downarrow_{s}$, pixelwise values decrease in the later stages.
This is evidence that features extracted in the later stages are better optimized for estimating $I^{SR}$ and $K$.
The decrease in value is noticeable in the edge lines of objects.
This also suggests that high-frequency features are represented more in the later stages than in the earlier stages; since the high-frequency features allow us to reconstruct the edge lines sharper, the residuals in the edge lines become smaller.

Note that pixelwise values in $R^{LR}_{t} = \tilde{I}^{LR}_{t} - I^{LR}$ are larger in all the stages (i.e., $t \in \{1,2,3,4\}$) because the LR loss is used only in the final output (i.e., $\tilde{I}^{LR} = (I^{SR} \ast K)\downarrow_{s}$) so that the final residual $R^{LR} = \tilde{I}^{LR} - I^{LR}$ is successfully reduced.
In KBPN, the LR loss is not used in all the stages for learning a variety of features in these stages; if $R^{LR}_{t}$ is converged to zero by the LR loss in each stage, $\tilde{I}^{LR}_{t}$ in all stages become similar.
Such uniform learning goals disturb all the stages to learn a variety of features.

\section{Limitation}
\label{section:limitation}

We have confirmed that KBPN exhibits performance that surpasses the SOTA methods in blind SR. However, our approach has the following limitations:

The number of parameters is substantial compared to conventional methods. Nevertheless, considering the recent trends in the development of foundation models~\cite{foundationmodel} and the advancements in computational resources, we believe that this is not a concern for offline applications. If our approach were to be introduced to edge devices, strategies such as pruning the dense connection, which can become a bottleneck in model size, are considered effective.

While KBPN demonstrates robustness against degradation caused by blur, it does not model degradation due to noise, thus not guaranteeing robustness in such scenarios. To address this issue, it is possible to train the model to handle various forms of degradation, such as the pure synthetic data generation technique~\cite{wang2021real} and the implicit degradation representation~\cite{xia2022knowledge}.

\section{Concluding Remarks}
\label{section:conclusion}

This paper proposed blind super-resolution methods, KCBPN and KBPN.
They have blur and SR branches.
Blur kernels estimated in the blur branch are employed not only for back-propagation using the kernel loss but also for SR feature enhancement in the SR branch.
In addition to the kernel loss, the LR loss is proposed to train KBPN with the estimated raw kernel.
Furthermore, KBPN achieves further SR feature enhancement using a residual between the input LR image and the one reconstructed by Eq.~(\ref{eq:down_model}).

Future work includes the improvement of the base SR and blur estimation networks.
%
%
Since only artificial blurs are used for training KCBPN and KBPN as well as other blind SR methods, difficulty in real-image SR is a possible limitation.
%
Further improvement may be achieved by integrating KCBPN and KBPN with a more variety of blurs~\cite{BSRGAN2021}.

This work is partly supported by JSPS KAKENHI Grant Numbers 19K12129 and 22H03618.

{\small
\bibliographystyle{IEEEtran}
\bibliography{v2}
}


\section{Biography Section}
\vspace{-2em}

\begin{IEEEbiography}[{\includegraphics[width=1in,height=1.25in,clip,keepaspectratio]{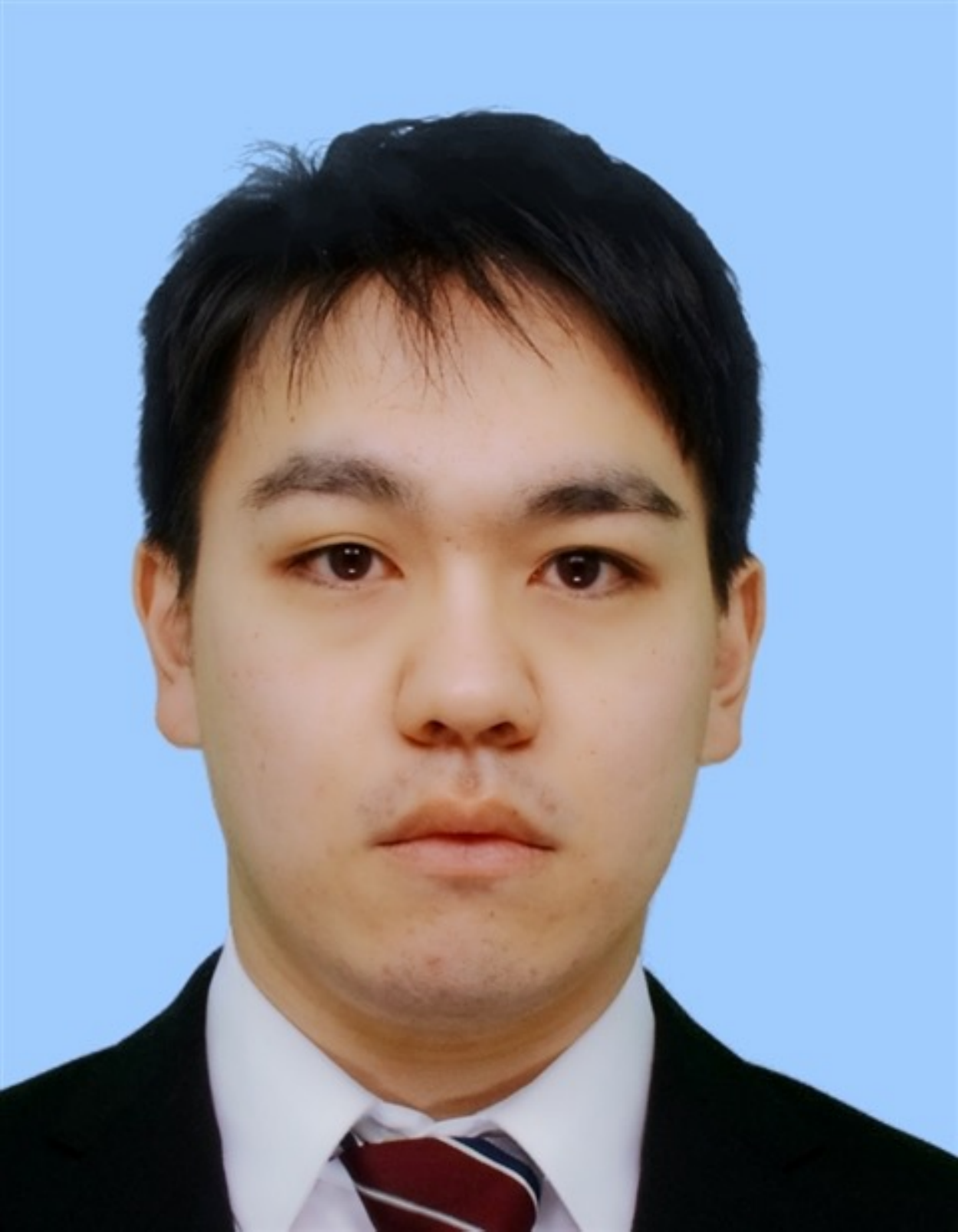}}]{Tomoki Yoshida} received the bachelor and master degrees in engineering from Toyota Technological Institute in 2019 and 2021, respectively. His research interests include single-image super-resolution and its enhancement for perceptual quality improvement and blind super-resolution. His award includes the winner award in NTIRE 2018 challenge.
\end{IEEEbiography}

\begin{IEEEbiography}[{\includegraphics[width=1in,height=1.25in,clip,keepaspectratio]{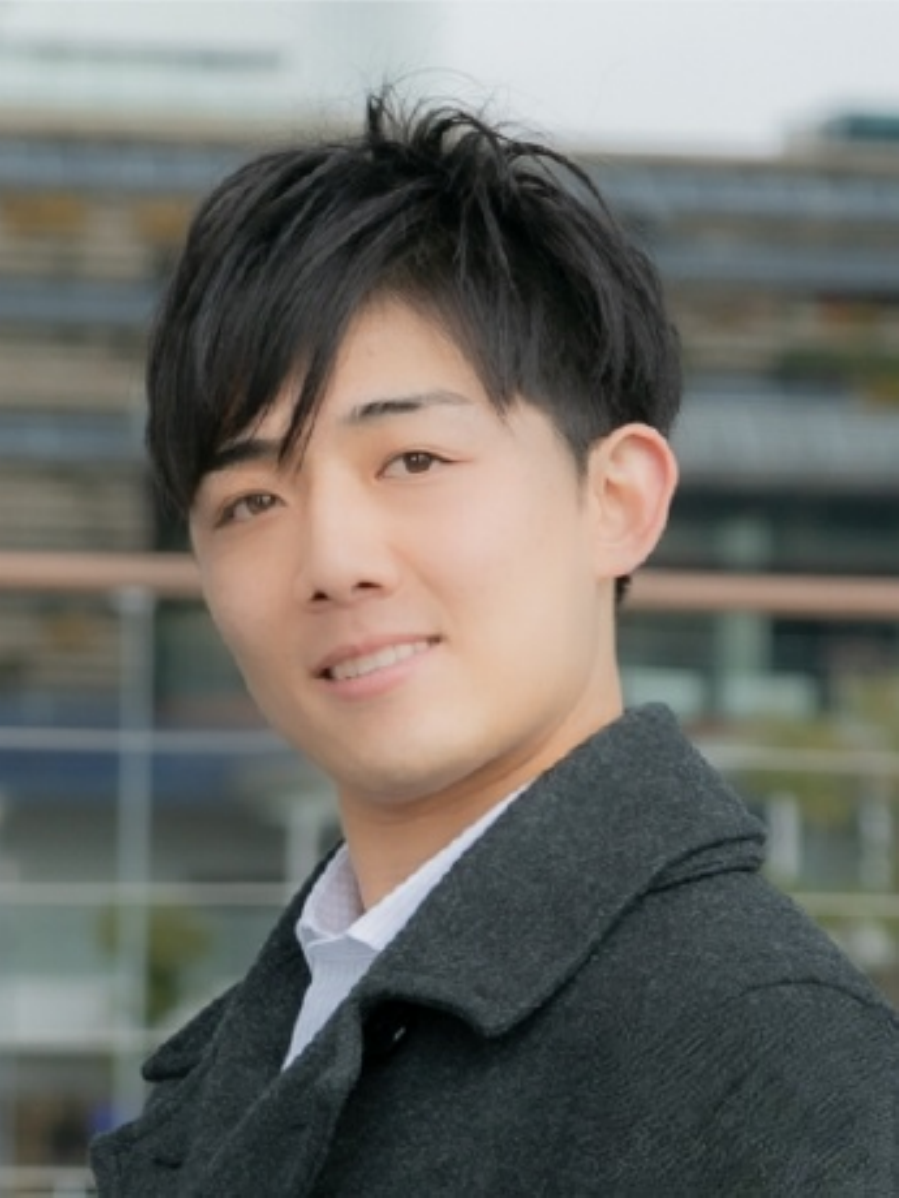}}]{Yuki Kondo} received the bachelor degree in engineering from Toyota Technological Institute in 2022. 
Currently, he is a researcher at Toyota Technological Institute. His research interests include low-level vision including image and video super-resolution and its application to tiny image analysis such as crack detection. His award includes the best practical paper award in MVA2021.
\end{IEEEbiography}

\begin{IEEEbiography}[{\includegraphics[width=1in,height=1.25in,clip,keepaspectratio]{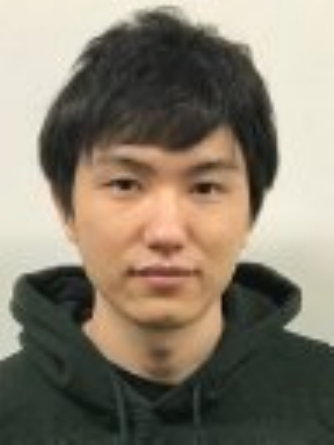}}]{Takahiro Maeda}
received the bachelor and master degrees in engineering from Toyota Technological Institute in 2019 and 2021, respectively. His research interests include low-level vision and human motion tracking, synthesis, and prediction. His awards include the best poster award in MVA2019 and the first place in Real Robot Challenge 2020, the Max Planck Institute for Intelligent Systems.
\end{IEEEbiography}

\begin{IEEEbiography}[{\includegraphics[width=1in,height=1.25in,clip,keepaspectratio]{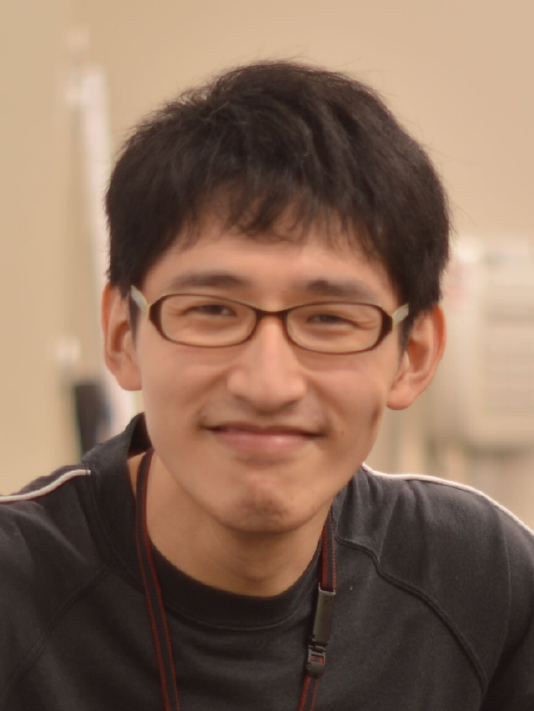}}]{Kazutoshi Akita}
received the bachelor and master degrees in engineering from Toyota Technological Institute in 2019 and 2021, respectively. His research interests include low-level vision including image and video super-resolution and its application to scale-free object detection. His awards include the winner award in NTIRE 2018 challenge on image super-resolution and 1st place in PIRM 2018 perceptual SR challenge.
\end{IEEEbiography}

\begin{IEEEbiography}[{\includegraphics[width=1in,height=1.25in,clip,keepaspectratio]{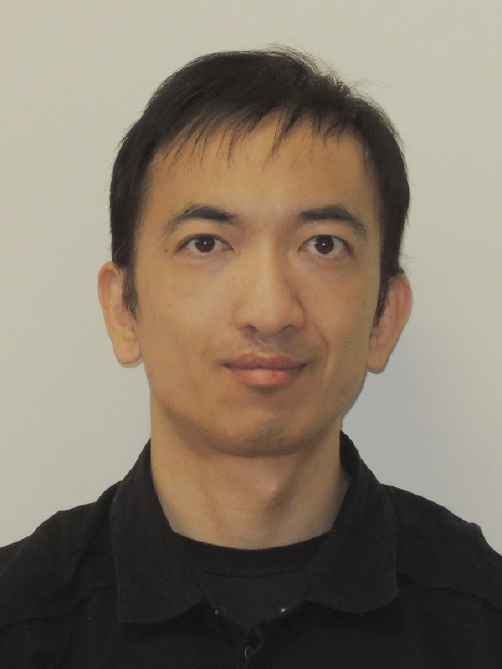}}]{Norimichi Ukita} received the B.E. and M.E. degrees in information engineering from Okayama University, Japan, in 1996 and 1998, respectively, and the Ph.D. degree in Informatics from Kyoto University, Japan, in 2001.
  From 2001 to 2016, he was an assistant professor (2001 to 2007) and an associate professor (2007-2016) with the graduate school of information science, Nara Institute of Science and Technology, Japan. In 2016, he became a professor at Toyota Technological Institute, Japan. He was a research scientist of Precursory Research for Embryonic Science and Technology, Japan Science and Technology Agency, during 2002--2006, and a visiting research scientist at Carnegie Mellon University during 2007--2009.
  Currently, he is also an adjunct professor at Toyota Technological Institute at Chicago.
  Prof. Ukita's awards include the excellent paper award of IEICE (1999), the winner award in NTIRE 2018 challenge on image super-resolution, 1st place in PIRM 2018 perceptual SR challenge, the best poster award in MVA2019, and the best practical paper award in MVA2021.
\end{IEEEbiography}

\end{document}